\documentclass[twocolumn, twocolappendix]{aastex631}

\usepackage{physics}
\usepackage[caption=false]{subfig}

\shorttitle{Synchrotron-Regulated Relativistic MHD Turbulence}
\shortauthors{X. Sun, L. Comisso, L. Sironi, A. Spitkovsky, A. Philippov}

\begin{document}

\title{Synchrotron-Regulated Relativistic Magnetohydrodynamic Turbulence:\\ Emission, Polarization, and Faraday Rotation}

\correspondingauthor{Xiaochen Sun}

\author[0009-0009-7676-6188]{Xiaochen Sun}
\affiliation{Department of Astrophysical Sciences, Princeton University, Princeton, NJ 08544, USA}
\affiliation{Department of Astronomy and Columbia Astrophysics Laboratory, Columbia University, New York, NY 10027, USA}
\email{xs2612@columbia.edu}

\author[0000-0001-8822-8031]{Luca Comisso}
\affiliation{Department of Astronomy and Columbia Astrophysics Laboratory, Columbia University, New York, NY 10027, USA}

\author[0000-0002-1227-2754]{Lorenzo Sironi}
\affiliation{Department of Astronomy and Columbia Astrophysics Laboratory, Columbia University, New York, NY 10027, USA}
\affiliation{Center for Computational Astrophysics, Flatiron Institute, 162 5th Avenue, New York, NY 10010, USA}

\author[0000-0001-9179-9054]{Anatoly Spitkovsky}
\affiliation{Department of Astrophysical Sciences, Princeton University, Princeton, NJ 08544, USA}

\author[0000-0001-7801-0362]{Alexander Philippov}
\affiliation{Department of Physics, University of Maryland, 7901 Regents Drive, College Park, MD 20742, USA}
\affiliation{Physics Department, Stanford University, 382 Via Pueblo Mall, Stanford, CA 94305, USA}
\affiliation{Kavli Institute for Particle Astrophysics and Cosmology, Physics and Astrophysics Building, 452 Lomita Mall, Stanford University, Stanford, CA 94305- 4085, USA}

\begin{abstract}
Relativistic magnetized plasmas in many high-energy astrophysical systems are both turbulent and strongly radiative, yet their nonlinear dynamics and radiative outcomes remain poorly understood. Here we present results from three-dimensional driven turbulence simulations in relativistic magnetohydrodynamics with synchrotron cooling. We compute Faraday rotation measures, synthetic synchrotron spectra and linear polarization maps from the simulated turbulence. The balance between energy injection from turbulent driving and synchrotron cooling keeps the plasma, on average, relativistically hot, thereby influencing the rotation measure. Synchrotron cooling triggers the thermal instability and drives the plasma into hot dilute and cold dense phases, which enhances the spatial and temporal variability of  synchrotron emission, especially at high frequencies. These diagnostics show qualitative similarities to observations of fast radio bursts, pulsar wind nebulae, and blazars, suggesting that turbulence may play an important role in shaping emission and propagation effects around high-energy sources.
\end{abstract}

\section{Introduction}
Relativistic magnetized turbulence is expected to be ubiquitous in high-energy astrophysical systems such as pulsar wind nebulae (PWNe), active galactic nuclei (AGN) jets, hot accretion flows, and gamma-ray burst (GRB) outflows \citep[e.g.,][]{2014ARA&A..52..529Y,2014MNRAS.438..278P,2015PhR...561....1K,2019ARA&A..57..467B}. Plasma flows around central engines, often magnetically dominated, are destabilized by instabilities, such as the kink instability \citep[e.g.,][]{1998ApJ...493..291B, 1999MNRAS.308.1006L, 2012ApJ...757...16M, 2019ApJ...884...39B}, the magneto-rotational instability \citep[e.g.,][]{1996ApJ...463..656S, 1998RvMP...70....1B, 2003ApJ...592.1060D}, the Kelvin-Helmholtz instability \citep[e.g.,][]{1961hhs..book.....C, 1984JGR....89..801M, 2013PhRvE..87d3101H, 2021ApJ...907L..44S} and the magnetic Rayleigh–Taylor instability \citep[e.g.,][]{1954RSPSA.223..348K, 1995ApJ...453..332J, 2012MNRAS.423.3083M, 2023PhRvR...5d3023Z}. The nonlinear evolution of these instabilities generically drives turbulence, leading to chaotic relativistic plasma motion and disordered magnetic fields. The resulting turbulence transfers energy from large to small scales, and ultimately heats the plasma to relativistic temperatures through dissipation.

The dissipated energy can be efficiently radiated away if the turbulent region is optically thin. In many such systems, this radiative output is dominated by synchrotron emission from relativistic electrons \citep[e.g.,][]{2010A&A...523A...2M}. Spectral modeling of individual sources indicates that the synchrotron luminosity can account for a few percent, and in some cases tens of percent, of the total power supplied by the central engine \citep[e.g.,][]{2000MNRAS.314..183W, 2012Sci...338.1445N, 2013ApJ...776...95F, 2014Sci...343...48M, 2014RPPh...77f6901B, 2014Natur.515..376G, 2008MNRAS.385..283C, 2020A&A...637A..86M, 2021MNRAS.504..878D}. Synchrotron cooling may therefore substantially influence plasma thermodynamics and turbulent dynamics.

Previous theoretical studies have suggested that plasma thermodynamics is regulated by the interplay between turbulence and synchrotron cooling: for a fixed total radiative power, the mean plasma temperature is set by the Thomson optical depth \citep{2018MNRAS.477.2849U}. Meanwhile, synchrotron cooling can trigger the thermal instability and lead to the formation of a multiphase medium \citep{1967ApJ...150..105S, 1979ApJ...233..463E}.  

Direct numerical simulations are essential to capture the interplay between turbulence and synchrotron cooling. Yet, the evolution of synchrotron-cooled relativistic turbulence remains unexplored in the magnetohydrodynamic (MHD) regime (see e.g., \citealp{2021PhRvL.127y5102C} for the collisionless kinetic regime).

Here we present the first relativistic MHD simulations of driven turbulence with self-consistent synchrotron cooling. Turbulence is sustained by a large-scale stirring force, with continuous energy injection. Such driven turbulence setups are standard in Newtonian (M)HD turbulence \citep[e.g.,][]{1988CF.....16..257E, 1998ApJ...508L..99S, 2001ApJ...554.1175M, 2002CoPhC.147..471B, 2003MNRAS.345..325C, 2010A&A...512A..81F, 2018ApJ...858L..19G}.  Driven turbulence and decaying turbulence exhibit similar inertial-range properties, supporting the use of driven turbulence simulations \citep[e.g.,][]{1998ApJ...508L..99S, 2009ApJ...691.1092L}.  For synchrotron cooling, we assume emission from electrons (and positrons) having an isotropic Maxwell–Jüttner distribution, as is commonly adopted in GRMHD simulations \citep[e.g.,][]{2009ApJ...706..497M, 2016ApJ...822...34P, 2016MNRAS.462..115D, 2021ApJ...910L..13E}. The balance between continuous energy injection and radiative losses establishes a quasi-steady turbulent state, in which synchrotron cooling self-consistently regulates the plasma thermodynamics.

We compute synthetic synchrotron spectra and polarization maps from the simulated turbulence, which may help us interpret the observational signatures from AGNs and PWNe. In high-synchrotron-peaked blazars such as Mrk~421 and Mrk~501, the polarized flux exhibits stochastic variability, commonly attributed to a turbulent environment  \citep[e.g.,][]{2008Natur.452..966M, 2010ApJ...716...30A, 2022Natur.611..677L, 2025ApJ...986..230M, 2025A&A...695A.217M, 2025A&A...703A..19C}. Spatial inhomogeneity and time variability of the Crab and Vela nebulae also support a turbulent magnetic-field structure \citep[e.g.,][]{2003ApJ...591.1157P, 2004ApJ...615..794B, 2022Natur.612..658X, 2023NatAs...7..602B}. 

Turbulence may also influence the propagation of polarized emission via Faraday rotation. In repeating fast radio burst (FRB) sources 20180301A and 20190303A, the polarization angle varies on timescales of days, which is attributed to a turbulent environment near the source \citep[e.g.,][]{2018Natur.553..182M,2022NatCo..13.4382W,2023Sci...380..599A,2025A&A...694A..75B,2025ApJ...979L..41L}. To facilitate comparison with observations, we compute Faraday rotation by integrating along the line of sight through the simulated turbulence, assuming a background polarized source.

This paper is organized as follows. Section~\ref{sec::numerical_setup} describes the numerical setup, including the driving force and the synchrotron cooling implementation. Section~\ref{sec::diagnostics} presents the emergence of hot-dilute/cold-dense thermodynamic phases, resulting from the interplay between turbulent stirring and synchrotron cooling, and analyzes the quasi-steady power spectral density of magnetic and kinetic fluctuations. Section~\ref{sec::prediction} derives observable signatures from the simulations: synchrotron spectra, polarization, and Faraday rotation measures. We discuss astronomical implications in Section~\ref{sec::application} and summarize our findings in Section~\ref{sec::conclusion}. Appendices provide details on the turbulence driving method, the numerical convergence study, the impact of the initial magnetization, and the thermal instability analysis.

\section{Numerical setup}
\label{sec::numerical_setup}
We implement the turbulence stirring force and the synchrotron cooling in the ideal special relativistic MHD module of \texttt{Athena++} \citep{2020ApJS..249....4S}, treating both the stirring force $\mathbb{F} = \left(\boldsymbol{F}, F_t\right)$ and the synchrotron cooling power $P_\text{syn}$ as source/sink terms in the momentum-energy equations \citep{2009ApJ...692..411N}:
\begin{align}
    &\partial_t \boldsymbol{M} + \nabla \cdot [ \left( \gamma^2 w \boldsymbol{u}\boldsymbol{u} \right) - \frac{\boldsymbol{B}\boldsymbol{B}}{ \gamma^2} - \left(\boldsymbol{u}\cdot \boldsymbol{B}\right)\left(\boldsymbol{u}\boldsymbol{B}+\boldsymbol{B}\boldsymbol{u}\right) \notag\\ 
    &\hspace{1cm}- \gamma^2 \left(\boldsymbol{u}\cdot \boldsymbol{B}\right)^2 \boldsymbol{u}\boldsymbol{u} + \left(P_g+P_m\right) \mathbb{I}] = \boldsymbol{F} - \gamma \boldsymbol{u} P_\text{syn} , \label{equ::mom_equ_drive}\\
    &\partial_t \varepsilon + \nabla \cdot \boldsymbol{M} = F_t - \gamma P_\text{syn}. \label{equ::energy_equ_drive}
\end{align} 
All variables follow standard definitions: fluid-frame gas density $\rho$, velocity $\boldsymbol{u}$, Lorentz factor $\gamma \equiv \left(1 - u^2 / c^2\right)^{-1/2}$, lab-frame magnetic field $\boldsymbol{B}$, total enthalpy density in the fluid frame $w \equiv w_g + 2 P_m$ with gas enthalpy density $w_g \equiv \rho c^2 + \frac{\Gamma}{\Gamma - 1} P_g$, adiabatic index $\Gamma$, gas pressure $P_g$, magnetic pressure in the fluid frame $P_m \equiv \left(B^2/\gamma^2 + \left(\boldsymbol{u} \cdot \boldsymbol{B}\right)^2 / c^2\right)/2$, total lab-frame momentum density $\boldsymbol{M} \equiv \gamma^2 \left(w - \left(\boldsymbol{u} \cdot \boldsymbol{B}\right)^2 / c^2\right) \boldsymbol{u} - \left(\boldsymbol{u} \cdot \boldsymbol{B}\right) \boldsymbol{B}$, and total lab-frame energy density $\varepsilon \equiv \gamma^2 w - \gamma^2 \left(\boldsymbol{u} \cdot \boldsymbol{B}\right)^2 / c^2 - \left(P_g + P_m\right)$. The magnetic permeability is unity, and the factor of $1 / \sqrt{4 \pi}$ is absorbed into $\boldsymbol{B}$. The mass conservation and induction equations, based on the frozen-in condition, remain unchanged, 
\begin{align}
    &\partial_t \left(\gamma \rho\right) + \nabla \cdot \left(\gamma \rho \boldsymbol{u}\right) = 0, \label{equ::mass_conservation}\\
    &\partial_t \boldsymbol{B} - \nabla \times \left(\boldsymbol{u} \times \boldsymbol{B} \right) = 0. \label{equ::induction_equation}
\end{align}

The stirring force $\mathbb{F}$ follows the same principles as in Newtonian (M)HD turbulence simulations \citep[e.g.,][]{2009ApJ...691.1092L, 2013MNRAS.436.1245F}, obeying two criteria:
\begin{enumerate}
    \item Momentum is injected only into the gas bulk motion ($\gamma^2 w_g \boldsymbol{u}$) and Poynting flux ($-\left(\boldsymbol{u} \times \boldsymbol{B}\right) \times \boldsymbol{B}/c^2$), without directly heating the plasma (i.e., without changing the temperature $P_g / \rho c^2$).
    \item The net momentum is conserved, $\left\langle \boldsymbol{M} \right\rangle_V = 0$ at each timestep, where $\left\langle \cdot \right\rangle_V$ denotes volume average.
\end{enumerate}
To satisfy these conditions, the spatial component of the stirring force $\boldsymbol{F}$ consists of two components:
\begin{equation}
    \boldsymbol{F} = \boldsymbol{F}_{OU} \left(\boldsymbol{x}, t\right) + \boldsymbol{f} \left(t\right). \label{equ::net_force}
\end{equation}
The Ornstein-Uhlenbeck (OU) process $\boldsymbol{F}_{OU}$ continuously injects momentum, while the uniform force $\boldsymbol{f}$ ensures zero net momentum integrated over the whole volume. As the total momentum density changes under $\boldsymbol{F}$, the gas velocity changes, and the lab-frame kinetic energy density ($\gamma^2 w_g$) and the electric-field energy density ($( \boldsymbol{u}\times\boldsymbol{B})^2/c^2$) increase accordingly by $F_t$. Thus, $F_t$ is constrained by $\left(\boldsymbol{F}, \rho, P_g, \boldsymbol{u}, \boldsymbol{B}\right)$; details and benchmark tests are provided in Appendix~\ref{sec::turb_driving}.

We prescribe the stirring force $\boldsymbol{F}_{OU}$ in the following way: Fourier-space amplitude scaling as $\left|\boldsymbol{F}_{OU}(\boldsymbol{k}, t)\right| \propto k^{-1}$ for wavelengths ranging from the simulation box size $L$ to $L/4$, oriented such that $\nabla \cdot \boldsymbol{F} = 0$. Temporally, the force de-correlates exponentially, $\left(\boldsymbol{F}_{OU}(\boldsymbol{k}, t) \cdot \boldsymbol{F}_{OU}(\boldsymbol{k}, t+\tau)\right) \propto \exp(-1.6 \tau c /L)$. Its magnitude is constrained via $\left\langle \boldsymbol{M} \cdot \boldsymbol{F}_{OU} \right\rangle_V = 1280 \rho_0^2 c^3 / L$, where $\rho_0$ is the initial mass density.

The synchrotron cooling power per unit volume in the fluid frame, $P_\text{syn}$, depends on the local electron (positron) number density and momentum distribution and on the fluid-frame magnetic field $\boldsymbol{B}'$. As the MHD framework does not evolve the electron (positron) distribution function, we assume an isotropic Maxwell-Jüttner distribution\footnote{In collisionless turbulence, electrons (positrons) may depart from thermal equilibrium \citep[e.g.,][]{2019ApJ...886..122C, 2019PhRvL.122e5101Z, 2020ApJ...893L...7W, 2022PhRvL.128g5101N, 2023ApJ...949...71Z}, and their temperature can differ from that of protons in a proton-electron plasma \citep[e.g.,][]{2023ApJ...944...24Z}, especially under strong cooling.} with dimensionless temperature $\Theta_t \equiv P_g / \rho c^2$ \citep[see Appendix~B.2 in][]{2022A&A...658A.100C}.
\begin{equation}
    P_\text{syn} = \eta_\text{syn} \frac{c}{L} P_g \frac{B'^2}{\rho_0 c^2} \frac{K_3 \left(\Theta_t^{-1} \right)}{K_2 \left(\Theta_t^{-1} \right)} \, . \label{equ::cooling}
\end{equation}
Here, $L$ denotes the simulation box size, which also corresponds to the largest eddy scale, and $K_n$ is the $n$th order modified Bessel function of the second kind.\footnote{Here we approximate $K_3(x^{-1})/K_2(x^{-1})$ as $4x+1$, which overestimates the exact value by only $\lesssim 15\%$ near $\Theta_t \sim 0.5$ and recovers the correct asymptotic limits for $\Theta_t \ll 1$ and $\Theta_t \gg 1$. This approximation simplifies and stabilizes the cooling term in the implicit–explicit (IMEX) numerical scheme, preventing  overcooling of the plasma internal energy.} The dimensionless cooling efficiency, $\eta_\text{syn}$, quantifies the sensitivity of the cooling to variations in $\Theta_t$, $B'$ and $P_g$. Physically, $\eta_\text{syn}\equiv 4 \left(n_e d_e^3\right)^{-1} \left(L/d_e\right)$ refers to the Thomson optical depth \citep{2018MNRAS.477.2849U}, related to the electron skin-depth $d_e$ and the number density of electrons (positrons) $n_e$ contributing to the mass density $\rho_0$. For simplicity, we assume that the radiative power is isotropic in the fluid co-moving frame, so the lab-frame cooling power is $\gamma P_\text{syn}$ and the lab-frame cooling drag force is $\gamma \boldsymbol{u} P_\text{syn}$ \citep{2009ApJ...692..411N}.

We initialize a uniform, static gas with mass density $\rho_0$ and pressure $P_{g} \left(t = 0\right) = 2\rho_0 c^2$, threaded by a uniform magnetic field $B_0 = \sqrt{\rho_0 c^2}$ along the $z$-direction in a cubic periodic box. Therefore, initially, $\Theta_t = 2$ and $\sigma \equiv B^2 / \rho c^2 = 1$. The adiabatic index is set to $\Gamma = 4/3$, assuming a relativistically hot gas.  
Numerical floor and ceiling values are: $\rho_\text{min} = 5 \times 10^{-3} \rho_0$, $\gamma_\text{max} = 50$, $P_{g,\text{min}} = 5 \times 10^{-5} \rho_0 c^2$, $\sigma_\text{max} \equiv \left(B'^2 / \rho c^2\right)_\text{max} = 2 \times 10^3$, and $\beta_\text{min} \equiv \left(2 P_g / B'^2\right)_\text{min} = 5 \times 10^{-6}$. We explore different cooling efficiencies $\eta_\text{syn} = 6.4 \times 10^{-2}, 10^{-3}, 10^{-4}$, spanning  strong- to weak-cooling cases. The numerical grid has a resolution of $1024$ cells in each direction. The properties of turbulence with different initial magnetization and resolution are described in Appendix~\ref{sec::mag}. Numerical units are set as follows: length in units of $L$, time in $L/c$, velocity in $c$, and magnetic field in $\sqrt{\rho_0} c$. We solve the MHD equations using a van-Leer time integrator, HLLD Riemann solver, and piecewise linear reconstruction.

\begin{figure}
    \centering
    \includegraphics[width=0.48\textwidth]{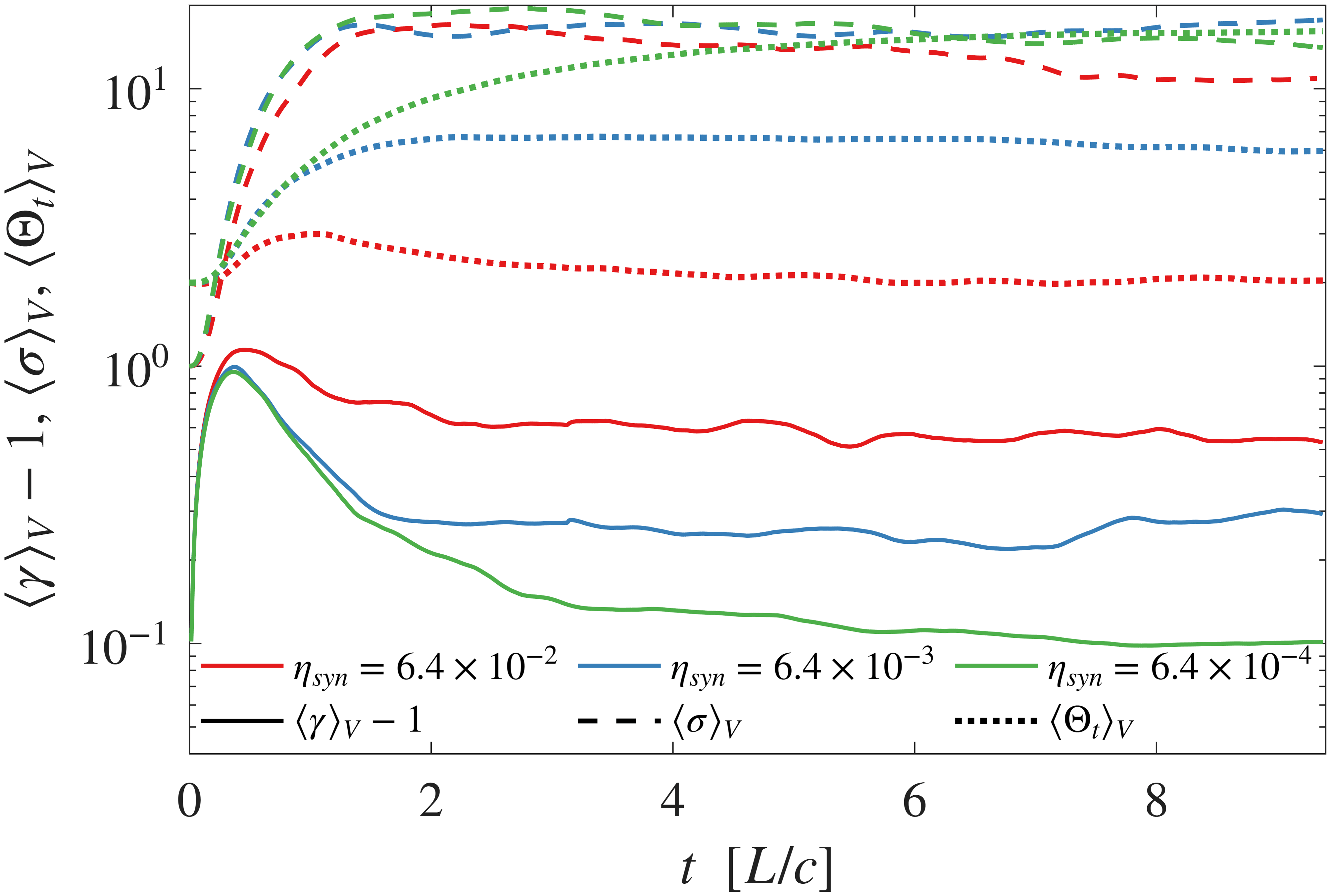}
    \caption{Time evolution of volume-averaged Lorentz factor $\left\langle \gamma \right\rangle_V$ (solid lines), magnetization $\left\langle \sigma \right\rangle_V$ (dashed lines), temperature $\left\langle\Theta_t \right\rangle_V$ (dotted lines). Colors indicate different cooling efficiencies as shown in the legend.}
    \label{fig::time}
\end{figure}
\label{sec::cooling_in_turb}
\begin{figure*}
    \centering
    \includegraphics[width=\textwidth]{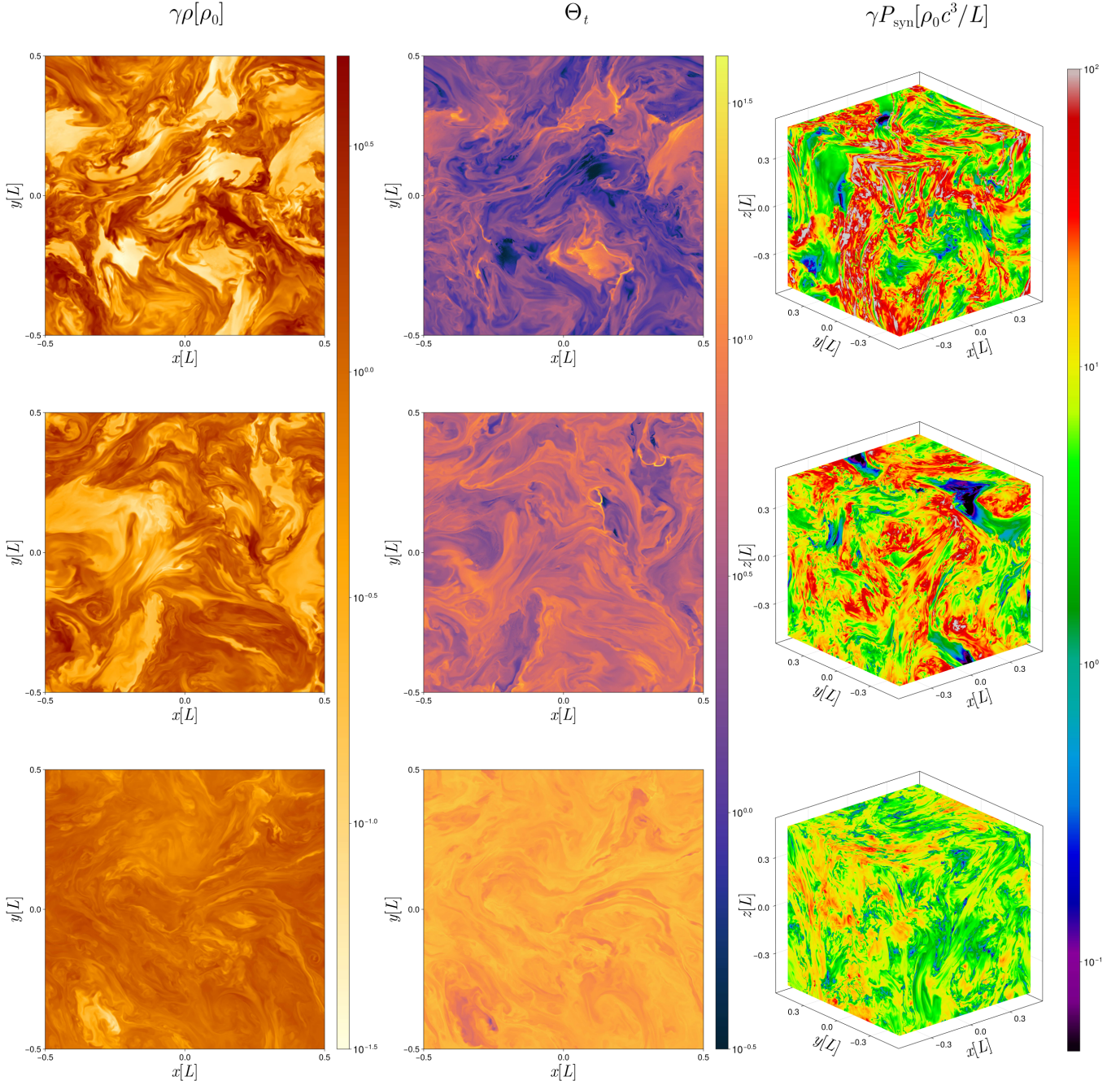}
    \caption{Snapshots of lab-frame density $\gamma \rho$ (left column), dimensionless temperature $\Theta_t$ (middle column), and synchrotron cooling power $\gamma P_\text{syn}$ (right column) at time $t c / L = 9.375$. From top to bottom, the rows represent different runs with $\eta_\text{syn} = 6.4 \times 10^{-2}, 10^{-3}, 10^{-4}$ respectively. The density and temperature slices are taken at $z = L/2$.}
    \label{fig::snapshot}
\end{figure*}
\section{Turbulence properties}
\label{sec::diagnostics}
Turbulence develops rapidly as energy is continuously injected by the stirring force. The injected energy first feeds the (electro-)magnetic fields and bulk plasma motions, driving sharp increases in the volume-averaged $\left\langle \sigma \right\rangle_V \equiv \left\langle B^2 \right\rangle_V /  \rho_0 c^2$ and $\left\langle \gamma \right\rangle_V \equiv \left\langle \gamma^2 \rho \right\rangle_V / \rho_0$ (see Figure~\ref{fig::time}). It then cascades to smaller scales and dissipates into heat, raising the volume-averaged temperature $\left\langle \Theta_t \right\rangle_V \equiv \left\langle \gamma P_g  \right\rangle_V /\rho_0 c^2$. As the gas heats up, synchrotron cooling gets stronger until the radiative power $\left\langle \gamma P_\text{syn} \right\rangle_V$ balances the energy injection rate $\left\langle F_t \right\rangle_V$ (see Appendix~\ref{sec::turb_driving}). As the internal energy reaches a statistical steady state, both the turbulent magnetic field and bulk motion settle into a quasi-steady state simultaneously.

In the following analysis, we focus on the quasi-steady state, after $\gtrsim 6 L/c$, during which the turbulence statistics remain approximately time-independent.
\subsection{Interplay of turbulence and synchrotron cooling}

Figure~\ref{fig::snapshot} displays quasi-steady turbulence snapshots of lab-frame density $\gamma \rho$, dimensionless temperature $\Theta_t$, and synchrotron cooling power $\gamma P_\text{syn}$. Because the energy injection rate $\left\langle F_t \right\rangle_V$ remains approximately constant across simulations (see Appendix~\ref{sec::turb_driving}), the total cooling power saturates at a similar level even as $\eta_\text{syn}$ varies by orders of magnitude. Consequently, to sustain a given cooling power, the highly cooling-efficient turbulence (i.e., higher $\eta_\text{syn}$) must saturate at a lower mean temperature, scaling as $\left\langle \Theta_t \right\rangle_V \propto \eta_\text{syn}^{-1/2}$ (Figure~\ref{fig::time}), consistent with \citet{2018MNRAS.477.2849U}. 

Beyond setting the mean temperature, the cooling efficiency also dictates the structure of density and temperature fluctuations. In cooling-efficient turbulence (i.e., $\eta_\text{syn}=6.4 \times 10^{-2}$), the plasma develops low-density, high-temperature phases to minimize radiative losses. To conserve the total mass $\left\langle \gamma \rho \right\rangle_V$, complementary high-density, low-temperature phases must emerge.

\begin{figure}
    \centering
    \includegraphics[width=0.45\textwidth]{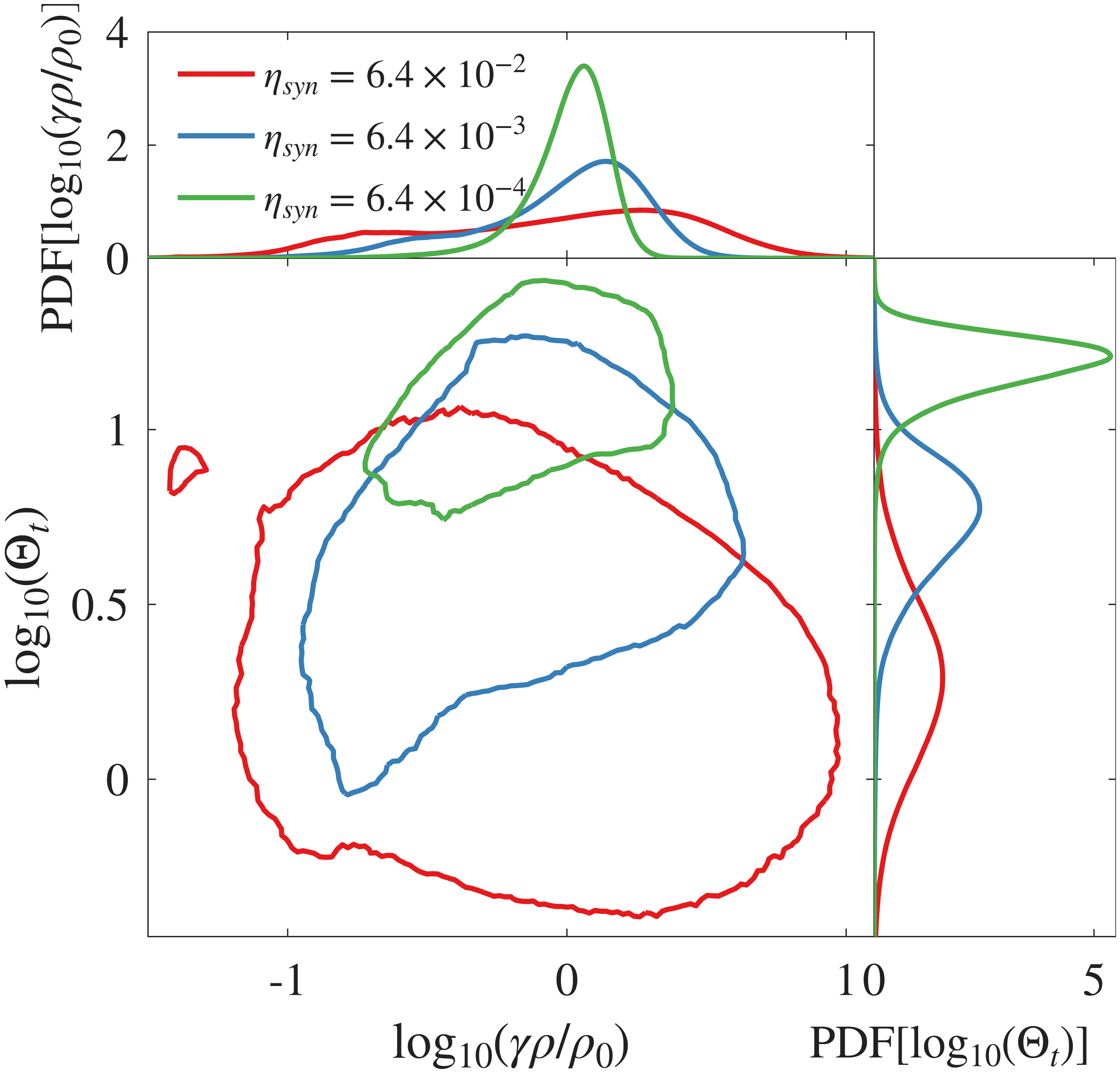}
    \caption{Lab-frame density $\gamma \rho$ versus temperature $\Theta_t$ distribution of quasi-steady turbulence, integrated from time $6\,L/c$ to $9.375 \,L/c$. The solid contours enclose $95\%$ of the plasma by volume. Colors indicate different cooling efficiencies as shown in the legend.}
    \label{fig::phase}
\end{figure}

The 2D distributions of the lab-frame density $\gamma \rho$ and temperature $\Theta_t$ in Figure~\ref{fig::phase} quantitatively illustrate how synchrotron cooling regulates the thermodynamics of the  turbulent plasma. As the cooling efficiency $\eta_\text{syn}$ increases, the plasma progressively bifurcates into distinct phases: a low density and high temperature phase, and a high density and low temperature phase. In the most cooling-efficient case, this bifurcation causes the lab-frame density distribution to display two peaks at $\sim \rho_0/5$ and $\sim 3 \rho_0$ (red line in the top panel of Figure~\ref{fig::phase}). Details are illustrated in Appendix~\ref{sec::phase}.

The emergence of distinct thermodynamic phases can be explained in terms of the synchrotron-cooling-induced thermal instability (\citealp{1967ApJ...150..105S, 1979ApJ...233..463E, 1992A&A...256..689B}; see also Appendix~\ref{sec::thermal_ins}). Over-dense regions carry stronger magnetic fields due to flux freezing. The stronger field increases the cooling losses (Equation~\ref{equ::cooling}), causing the local thermal pressure to drop. The surrounding plasma and magnetic fields are drawn into these collapsing regions. The positive feedback loop amplifies the initial density and temperature contrasts, ultimately separating the plasma into cold-dense and hot-dilute phases (Figure~\ref{fig::cool_instability_single} in Appendix~\ref{sec::thermal_ins}), consistent with Figure~\ref{fig::phase} as $\eta_\text{syn}$ increases. 

Meanwhile, turbulent mixing suppresses the runaway instability. The mixing timescale remains of order the eddy turnover time $L/c$, whereas the thermal-instability timescale decreases with increasing cooling efficiency, $P_g/P_\text{syn} \propto \eta_\text{syn}^{-1/2}$ \citep{2018MNRAS.477.2849U}. In the cooling-inefficient run ($\eta_\text{syn} = 6.4 \times 10^{-4}$), density and temperature remain nearly uniform (Figure~\ref{fig::snapshot}) and the phase distribution closely follows the adiabatic relation $\Theta_t \propto (\gamma \rho)^{\Gamma - 1}$ (Appendix~\ref{sec::phase}). As $\eta_\text{syn}$ increases, the phase distribution broadens as a result of the thermal instability; in the limit of extremely efficient cooling ($\eta_\text{syn} \gg 6.4 \times 10^{-2}$), clearly bimodal distributions may emerge in both $\gamma \rho$ and $\Theta_t$. The role of turbulent mixing in mitigating the impact of the thermal instability is supported by the statistics of density fluctuations, whose anisotropy scaling resembles the one expected for uncooled turbulence (Appendix~\ref{sec::struct_func}) rather than a thermal-instability pattern (Figure~\ref{fig::cool_instability_single} in Appendix~\ref{sec::thermal_ins}).

\subsection{Phase identification}
\begin{figure}
    \centering
    \includegraphics[width=0.45\textwidth]{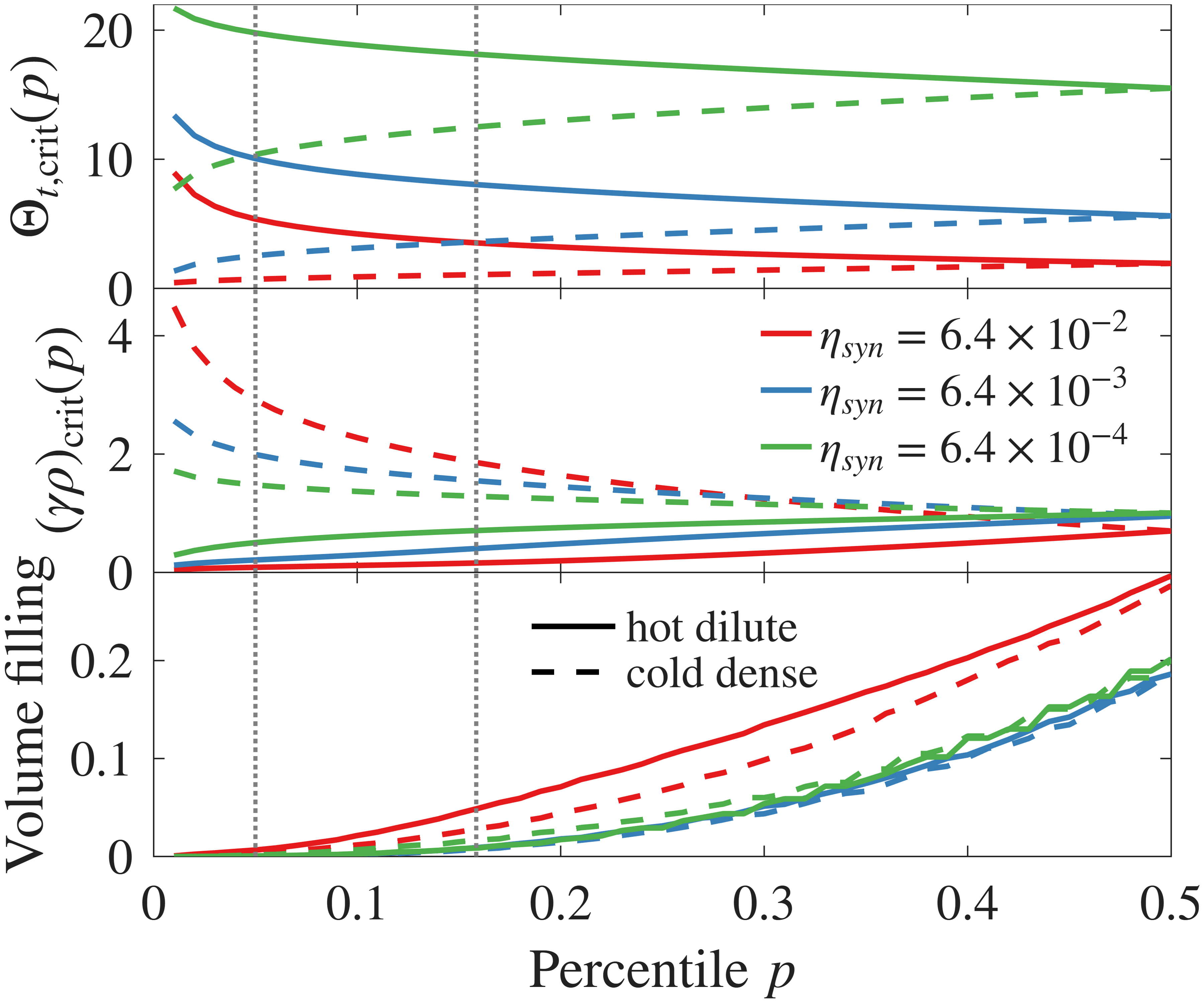}
    \caption{Temperature $\Theta_{t,\mathrm{crit}}(p)$ (top) and lab-frame density $(\gamma\rho)_{\mathrm{crit}}(p)$ (middle) thresholds as functions of the percentile $p$, together with the volume filling fractions (bottom) of the hot-dilute (solid) and cold-dense (dashed) phases. Colors indicate different cooling efficiencies as shown in the legend. The two vertical dotted lines mark $p=0.05$ and $0.158$, used to respectively identify the extreme and moderate hot-dilute and cold-dense phases in Figure~\ref{fig::phase_identification_snapshot}.}
    \label{fig::phase_identification}
\end{figure}
\begin{figure*}
    \centering
    \includegraphics[width=\textwidth]{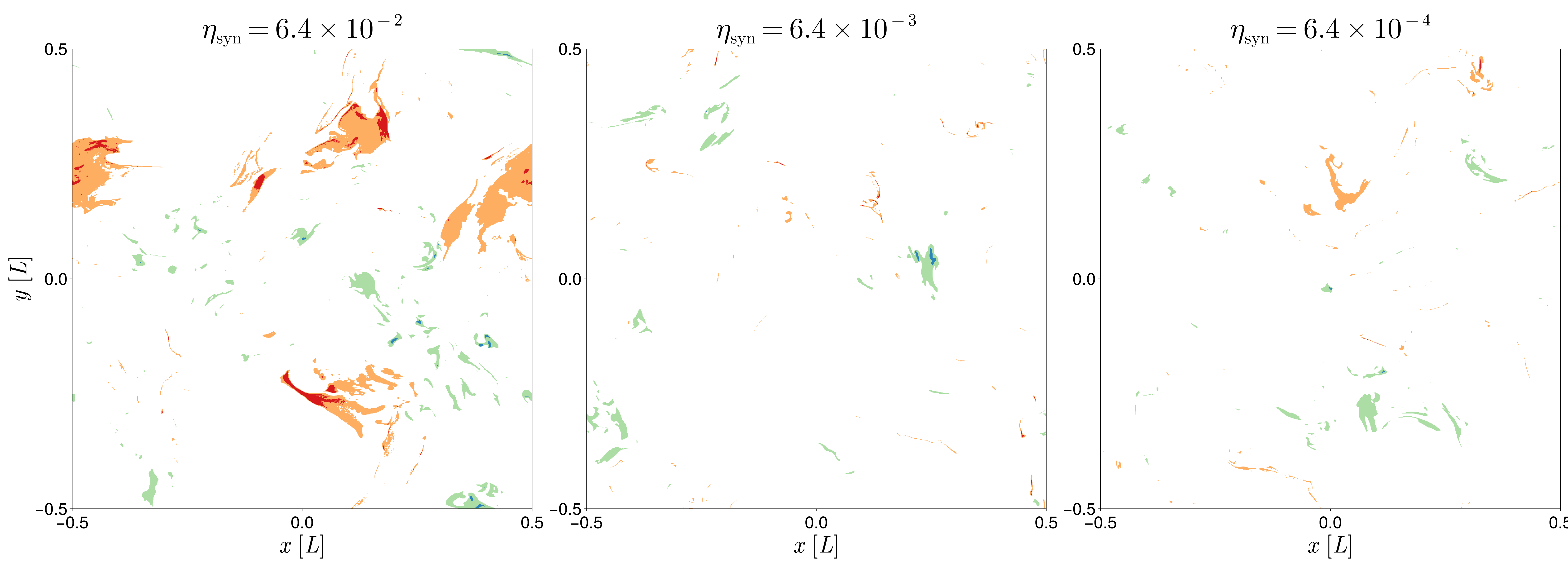}
    \caption{Snapshots of the identified hot-dilute and cold-dense phases in quasi-steady state. The extreme phases identified with $p=0.05$ are shown in red (hot-dilute) and blue (cold-dense), while the moderate phases identified with $p=0.158$ are shown in orange (hot-dilute) and green (cold-dense). All slices are taken at $z=L/2$ and $tc/L=9.375$, as in Figure~\ref{fig::snapshot}.}
    \label{fig::phase_identification_snapshot}
\end{figure*}
Here we identify the hot-dilute and cold-dense phases resulting from the synchrotron thermal instability, using thresholds in temperature, $\Theta_{t,\text{crit}}$, and lab-frame density, $(\gamma\rho)_\text{crit}$, motivated by the behavior of the instability (Figure~\ref{fig::cool_instability_single} in Appendix~\ref{sec::thermal_ins}).
The threshold values are determined by a given percentile $p$ in the 1D probability density functions (PDFs) of $\Theta_t$ and $\gamma\rho$ (Figure~\ref{fig::phase}). The hot-phase threshold $\Theta_{t,\mathrm{crit}}(p)$ corresponds to the hottest $p\%$ of the plasma volume, while the dilute-phase threshold $(\gamma\rho)_{\mathrm{crit}}(p)$ corresponds to the most dilute $p\%$; an analogous definition applies to the cold-dense phase. The threshold values depend on both the percentile $p$ and the specific run under consideration (Figure~\ref{fig::phase_identification}).

We combine the temperature and density thresholds to identify the hot-dilute and cold-dense phases, and compute their corresponding volume filling fractions in the bottom panel of Figure~\ref{fig::phase_identification}. The hot-dilute phase is defined by $\Theta_t>\Theta_{t,\mathrm{crit}}(p)$ and simultaneously $\gamma\rho<(\gamma\rho)_{\mathrm{crit}}(p)$, with an analogous definition for the cold-dense phase. In the case of pure thermal instability (Figure~\ref{fig::cool_instability_single} in Appendix~\ref{sec::thermal_ins}), the entire  plasma should reside in either the hot-dilute or cold-dense phase, so the volume filling fraction should be equal to the chosen percentile $p$. By contrast, purely adiabatic turbulence produces a positive correlation between $\Theta_t$ and $\gamma\rho$, and therefore no hot-dilute or cold-dense phases.

All our simulations exhibit non-zero volume filling fractions for both hot-dilute and cold-dense phases, confirming the presence of the synchrotron-cooling-induced thermal instability. The cooling-efficient run ($\eta_\text{syn}=6.4\times10^{-2}$) yields the largest filling fractions, whereas the two less efficient runs ($\eta_\text{syn}=6.4\times10^{-3}$ and $6.4\times10^{-4}$) produce nearly identical values. Nevertheless, for all runs the filling fraction remains below the threshold percentile $p$, as a result of turbulent mixing.

Figure~\ref{fig::phase_identification_snapshot} illustrates the identified hot-dilute and cold-dense phases using percentile thresholds of $p=0.05$ and $0.158$. The more extreme phases ($p=0.05$) are shown in red (hot-dilute) and blue (cold-dense), while the less extreme phases ($p=0.158$) are shown in orange (hot-dilute) and green (cold-dense). In all runs,
In all runs, both phases appear as randomly distributed, isolated regions embedded within a pervasive intermediate phase.
Unlike the sharp phase separation observed in the nonlinear stages of the pure thermal instability (Figure~\ref{fig::cool_instability_single} in Appendix~\ref{sec::thermal_ins}), this morphology suggests that turbulent mixing suppresses the full development of the instability. The cooling-efficient run ($\eta_\text{syn}=6.4\times10^{-2}$) produces both larger filling fractions and also more extended hot-dilute and cold-dense structures than the weaker-cooling runs. A more quantitative characterization of the phase morphology and topology, e.g., via the Euler characteristic or genus \citep[e.g.,][]{2007ApJ...658..423K, 2018MNRAS.475.1843M, 2026arXiv260413154C}, may be a promising direction for future work. The fractal dimension, estimated from box-counting \citep[e.g.,][]{2009ApJ...692..364F}, appears to be sensitive to the adopted threshold and therefore less informative, and it is not reported here.

\subsection{Turbulence power spectra}
\begin{figure}
    \centering
    \includegraphics[width=0.45\textwidth]{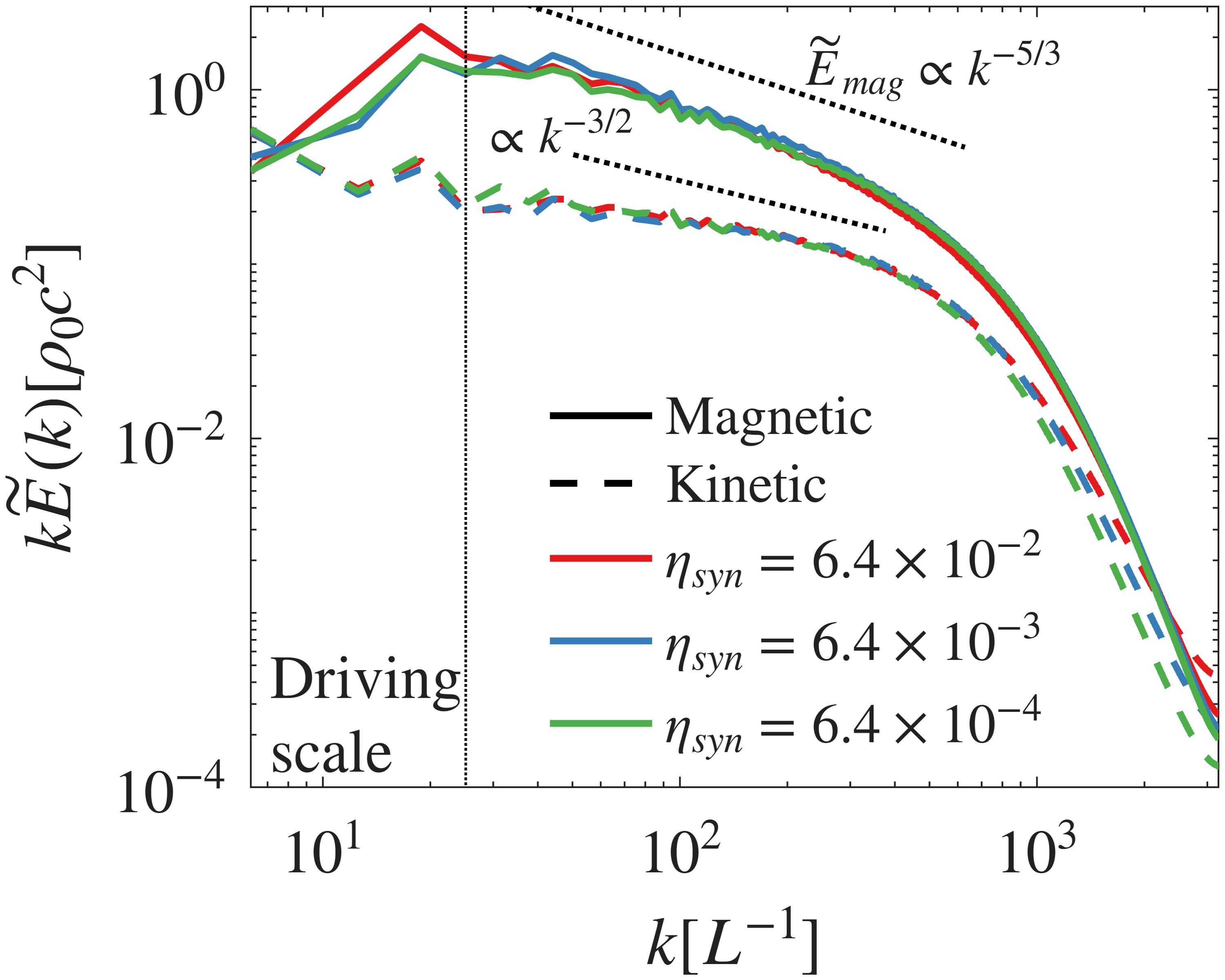}
    \caption{The shell-integrated magnetic and kinetic power spectral density $\tilde{E}(k)$ of quasi-steady turbulence . Colors indicate different cooling efficiencies as shown in the legend. Solid lines represent the magnetic power spectra $\tilde{E}_\text{mag}(k)$, while dashed lines refer to the kinetic power spectra $\tilde{E}_\text{kin}(k)$ (Equation~\ref{equ::kin_pwr_def}). The dotted lines represent the scalings $\tilde{E}(k) \propto k^{-5/3}$ and $\tilde{E}(k) \propto k^{-3/2}$, as labeled in the figure.}
    \label{fig::pwr}
\end{figure}
We decompose the magnetic and kinetic fluctuations in Fourier space to compute the power spectral density (PSD; Figure~\ref{fig::pwr}). The shell-integrated  PSD are computed as
\begin{align}
    \tilde{E}_\text{kin}(k) &= \int \dd \Omega_k k^2  \left|\tilde{\boldsymbol{\mathcal{M}}}_\text{kin} \left(k , \Omega_k \right)\right|^2 , \notag \\
    \tilde{E}_\text{mag}(k) &= \int \dd \Omega_k k^2  \left|\tilde{\boldsymbol{B}} \left(k , \Omega_k \right)\right|^2 / 2 , \notag \\
    \text{where }& \notag \\
    \tilde{\boldsymbol{\mathcal{M}}}_\text{kin} \left(\boldsymbol{k}\right) &\equiv \int \dd^3 \boldsymbol{x} \exp\left(-i \boldsymbol{k} \cdot \boldsymbol{x}\right) \sqrt{\frac{  w_{g,\text{lab}} \left(\boldsymbol{x}\right)}{\gamma \left(\boldsymbol{x}\right) + 1} } \gamma \boldsymbol{u}\left(\boldsymbol{x}\right), \notag \\
    \tilde{\boldsymbol{B}} \left(\boldsymbol{k}\right) &\equiv \int \dd^3 \boldsymbol{x} \exp\left(-i \boldsymbol{k} \cdot \boldsymbol{x}\right) \boldsymbol{B}\left(\boldsymbol{x}\right),\label{equ::kin_pwr_def}
\end{align}
with $\Omega_k$ denoting the solid angle in Fourier space. With this definition, the kinetic energy in configuration space, $E_\text{kin}(\boldsymbol{x})  \equiv \boldsymbol{\mathcal{M}}^2_\text{kin}\left(\boldsymbol{x}\right) = \left(\gamma - 1\right) w_{g,\text{lab}}$ recovers the relativistic expression of the kinetic energy density.\footnote{The PSD calculation largely follows the Newtonian treatment \citep[e.g.,][]{2009ApJ...691.1092L}, with a relativistic correction to the kinetic energy density. \citet{2023PhRvL.131e5201S} extends the kinetic PSD for cold plasmas moving at relativistic bulk speeds using $\boldsymbol{\mathcal{M}}_\text{kin} \equiv \sqrt{\gamma \rho/(\gamma+1)}\gamma\boldsymbol{u}$. For relativistically hot plasmas, as in our runs, the lab-frame mass density $\gamma \rho$ should be replaced by the enthalpy density $w_{g,\text{lab}}=\gamma w_g$ as a proxy of the fluid inertia.}

Turbulence with different cooling efficiencies exhibits broadly similar power spectra. In the inertial range ($4 \lesssim kL/(2\pi) \lesssim 96$), the magnetic PSD $\tilde{E}_\text{mag}(k)$ approximately follows a Kolmogorov-like scaling, $\tilde{E}(k) \propto k^{-5/3}$, whereas the kinetic PSD $\tilde{E}_\text{kin}(k)$ is slightly harder than $k^{-3/2}$. Magnetic fluctuations dominate over kinetic ones throughout the inertial range, while they become comparable around the dissipation range.

The similarity in PSD across cooling efficiencies indicates that the turbulent cascade is largely insensitive to the plasma thermodynamic state \citep[e.g.,][]{2020ApJ...889...19G}, especially when magnetic pressure dominates over gas pressure, as in our runs. Consistently, different $\eta_\text{syn}$ produce similar structure functions and anisotropy scalings (Appendix~\ref{sec::struct_func}). The amplitudes of magnetic and kinetic fluctuations are set by the large-scale eddies, primarily controlled by the stirring force and the total radiated power, which are similar among all runs. A Kolmogorov-like magnetic energy cascade appears to be a generic feature of relativistic MHD/force-free turbulence \citep[e.g.,][]{2005ApJ...621..324C,2012ApJ...744...32Z, 2016ApJ...817...89Z, 2021ApJ...923L..13C}. In contrast, the harder kinetic PSD likely reflects relativistic compressive modes with strong density/pressure fluctuations in the inertial range. A detailed analysis of the kinetic energy PSD is provided in Appendix~\ref{sec::kin}.

\section{Observational Signatures} 
\label{sec::prediction}
Relativistic turbulence has been linked to synchrotron emission from blazars and PWNe, as well as Faraday rotation in FRB sources. In this section, we compute polarized synchrotron emission and Faraday rotation from our simulated turbulence, to make a qualitative connection with these observational signatures.

\subsection{Polarized emission}
\label{sec::emission}
By assuming that electrons follow an isotropic Maxwell-Jüttner distribution with temperature $\Theta_t$ and  radiate in the rest-frame field $\boldsymbol{B}'$ (Section~\ref{sec::numerical_setup}), the synchrotron emissivity at frequency $\nu$ in the fluid rest frame can be approximated as
(\citealp{1996ApJ...465..327M, 2016ApJ...822...34P, 2022ApJS..262...28W, 2023ApJ...957..103G}; see also \citealp{2016MNRAS.462..115D} for an alternative fitting formula):
\begin{align}
j'_s \left(\nu\right) = & \frac{3 \times 2^{-29/6} }{1 + 2^{49/12}+5 \times 2^{1/6}} \eta_\text{syn} \frac{c \Omega_{e0}}{L} \times \notag \\ & \rho c^2 \frac{B' \sin \theta}{\sqrt{\rho_{0} c^2}}   \exp\left(-X^{1/3}\right) \times \notag \\
&\begin{cases}
\displaystyle \left(X^{1/2}+2^{11/12}X^{1/6}\right)^2, \;\text{s = (Stokes $I$)}\\
\displaystyle -\left(X^{1/2}+2^{11/12}\dfrac{7\Theta_t^{24/25}+35}{10 \Theta_t^{24/25}+75}\,X^{1/6}\right)^2, \\ \text{s = (Stokes $Q$)}
\end{cases} \label{equ::emissivity_fit} \\
&\text{where } X \equiv \frac{\nu / \left( 2 \pi \Omega_{e0} B'/\sqrt{\rho_{0} c^2}\right)}{  2/9 \Theta_t^2 \sin \theta }. \notag
\end{align}
Here, $\theta$ denotes the angle between $\boldsymbol{B}'$ and the photon propagation direction, and the emission frequency $\nu$ is normalized by the local electron cyclotron frequency $\Omega_e \equiv \Omega_{e0} B'/\sqrt{\rho_0 c^2} = e B'/(m_e c)$, where $\Omega_{e0}$ is defined for a background field strength equal to $\sqrt{\rho_0 c^2}$. The angle-integrated total emissivity, $\int \dd \nu \int 2\pi \sin\theta \dd\theta j'_I$, recovers the synchrotron cooling power $P_\text{syn}$ for $\Theta_t \gtrsim 1$. The negative sign of $j'_Q$ indicates that the electric vector is perpendicular to $\boldsymbol{B}'$ in the fluid frame. The circular polarization degree is negligibly small for synchrotron emission and is therefore not included in our analysis.

The Stokes parameters $(I, Q, U)$ are obtained by integrating the local emissivities along the line of sight (LoS) \citep[e.g.,][]{2011MNRAS.410.1052S,2011MNRAS.416.2574H}:
\begin{align}
    \frac{\dd I_\nu }{\dd s} =& j'_I\left(\frac{\nu}{\mathcal{D}}, \theta, B', \Theta_t, \rho\right) \times \mathcal{D}^2, \notag \\
    \frac{\dd Q_\nu }{\dd s} =& j'_Q \times \cos \left(2 \chi \right) \times \mathcal{D}^2, \notag \\
    \frac{\dd U_\nu }{\dd s} =& j'_Q \times \sin \left(2 \chi \right) \times \mathcal{D}^2.
\end{align}
Here $\mathcal{D}$ is the Doppler factor relating the fluid frame to the observer frame. We account for Doppler shifting and relativistic beaming via $\nu \to \nu/\mathcal{D}$ and $j'_s \to j'_s \mathcal{D}^2$. Absorption and Faraday effects are neglected in this subsection.
We also neglect light-travel-time delays and integrate directly over instantaneous turbulence snapshots. The angles $\theta$ and $\chi$ denote the local magnetic-field orientation in the observer coordinate frame.

We consider two different LoS:
\begin{enumerate}
\item \textbf{Observer perpendicular to the mean magnetic field:}
The line of sight is along the $x$-direction. The field-viewing angle is $\theta = \arccos(B'_x/B')$, and the Doppler factor is $\mathcal{D} = \gamma(1 + u_x/c) $. Linear polarization is defined by $Q = \langle \varepsilon_y^2 - \varepsilon_z^2\rangle$ and $U = \langle 2\varepsilon_y\varepsilon_z\rangle$, with $\boldsymbol{\varepsilon}$ the electric-field vector of the synchrotron radiation. The linear polarization angle in the lab frame is $\chi = \arctan(B'_z/B'_y)$.
\item \textbf{Observer parallel to the mean magnetic field:}
The line of sight is along the $z$-direction. Thus $\theta = \arccos(B'_z/B')$, $\mathcal{D} = \gamma(1 + u_z/c) $, and $\chi = \arctan(B'_y/B'_x)$.
\end{enumerate}
We classify the possible observational signatures according to whether the observer can spatially or temporally resolve the turbulent region.

\subsubsection{Spectra and linear polarization}
\begin{figure}
    \centering
    \includegraphics[width=0.45\textwidth]{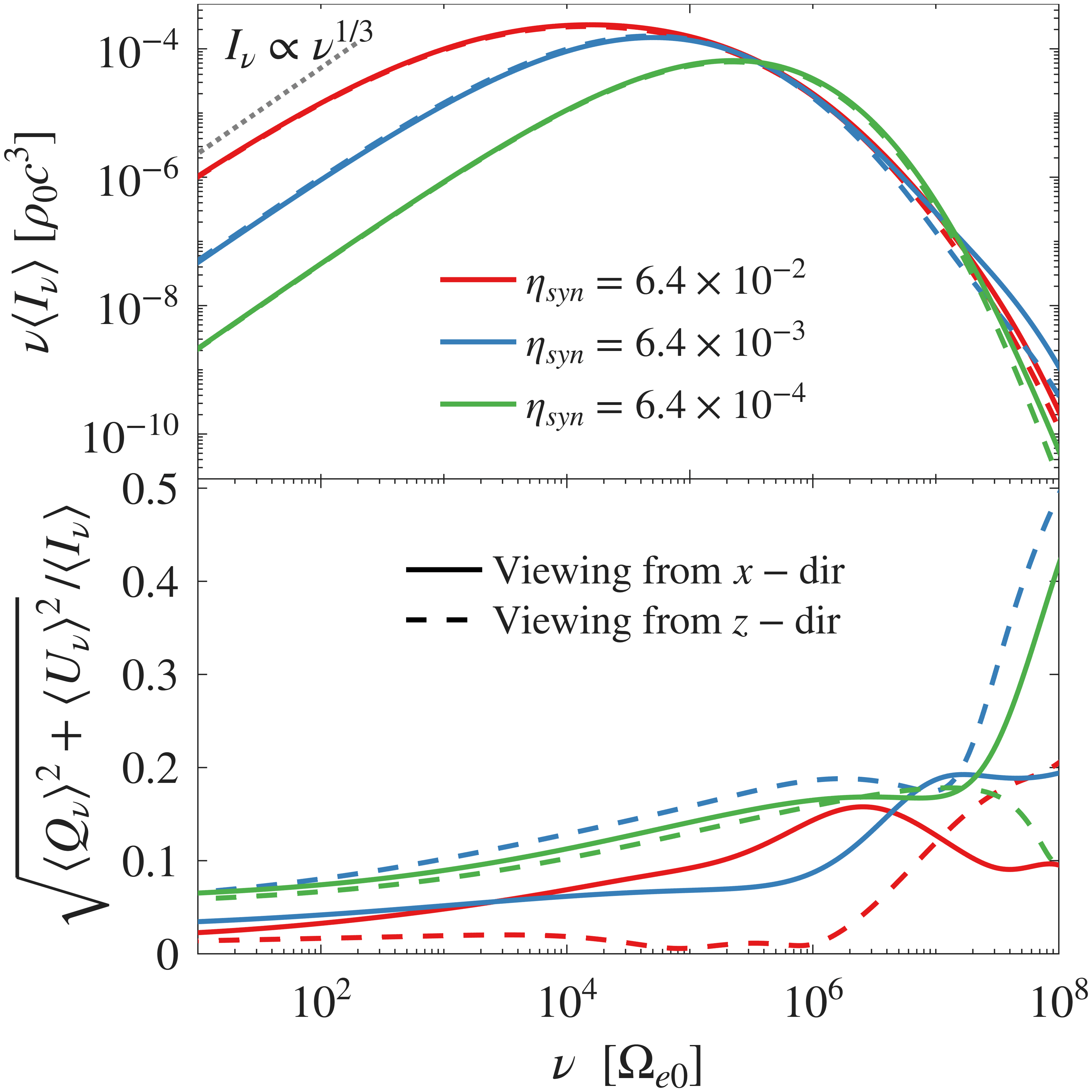}
    \caption{The synchrotron emission spectra ($\left\langle I_\nu\right\rangle$, top) and linear polarization degree ($\sqrt{\left\langle Q_\nu \right\rangle^2 + \left\langle U_\nu \right\rangle^2} / \left\langle I_\nu\right\rangle$, bottom) of quasi-steady turbulence. The emission frequency $\nu$ is in units of electron cyclotron frequency $\Omega_{e0}$, defined for the background field intensity equal to $\sqrt{\rho_0 c^2}$, and the angle brackets denote spatial and temporal averaging. Solid lines are for the observer perpendicular to the mean magnetic field direction, while dashed lines are for the observer along the mean magnetic field direction. Colors indicate different cooling efficiencies as shown in the legend.}
    \label{fig::emission_total}
\end{figure}
In realistic observing conditions, the measured emission is typically an integration over many turbulent eddies and many turnover times. We therefore first examine the synchrotron spectrum from the entire  turbulence box in quasi-steady state (Figure~\ref{fig::emission_total}), averaged over the whole volume and over time (after $6.25 L/c$).

The total synchrotron intensity $\langle I_\nu \rangle$ exhibits a broad spectrum because the plasma temperature $\Theta_t$ varies widely in the turbulent volume (Figure~\ref{fig::phase}), especially for the most strongly cooled case. At low frequencies, emission arises from the entire plasma, with each cell contributing a $\nu^{1/3}$ spectrum below its peak frequency (Equation~\ref{equ::emissivity_fit}). At higher frequencies, the emission of each cell cuts off exponentially ($X \gtrsim 1$), but the broad temperature distribution leads to a superposition of many of such cutoffs, producing an extended spectrum. Cooling-efficient turbulence ($\eta_\text{syn} = 6.4 \times 10^{-2}$) yields the lowest peak frequency \citep{2018MNRAS.477.2849U} and the broadest spectral width, since it displays the widest  range of temperatures (Figure~\ref{fig::phase}). Variations in magnetic field strength, density, and Doppler boosting play secondary roles. The spectra also  depend only marginally on the LoS direction, compare solid versus dashed lines in Figure~\ref{fig::emission_total}.

The linear polarization degree, $\sqrt{\langle Q_\nu\rangle^2+\langle U_\nu\rangle^2}/\langle I_\nu\rangle$, remains at the $\sim 5-10\%$ level at low and intermediate frequencies, due to strong depolarization from randomly oriented magnetic fields when integrating along the LoS \citep{2002ApJ...573..485R}. As the turbulent magnetic field is statistically isotropic and much stronger than the mean field (Figure~\ref{fig::time}), the net polarization is much lower than the theoretical maximum expected for an ordered field. At higher frequencies, the emission is dominated by rare, hot plasma columns with strong magnetic fields. There, the effect of LoS depolarization is reduced and the polarization degree can reach $\gtrsim 50\%$. The spatial and temporal intermittency of these hot  columns implies that the resulting polarized signal can vary sharply with viewing angle and time.

\subsubsection{Images and statistics: spatial/temporal variability}
\label{sec::syn_stat}
\begin{figure*} 
    \centering
    \includegraphics[width=\textwidth]{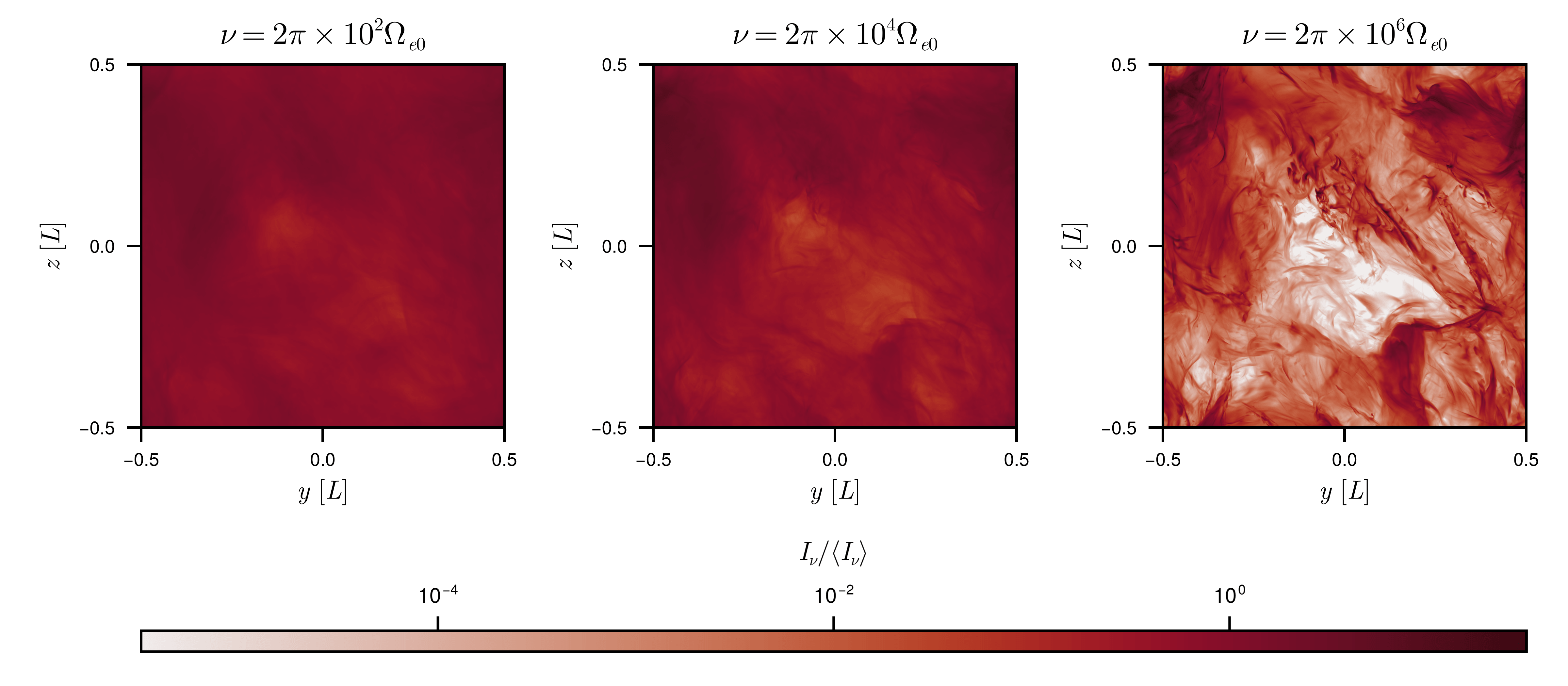}
    \caption{Synchrotron intensity maps of $I_\nu / \langle I_\nu \rangle$ from turbulence with $\eta_\text{syn} = 6.4 \times 10^{-3}$, viewed along the $x$ direction at $t c / L = 9.375$, normalized by the spatial and temporal average at the corresponding frequency $\langle I_\nu \rangle$ (Figure~\ref{fig::emission_total}). Panels from left to right show emission from low to high frequencies, as marked on the plot.}
    \label{fig::syn_image}
\end{figure*}
\begin{figure}
    \centering
    \includegraphics[width=0.45\textwidth]{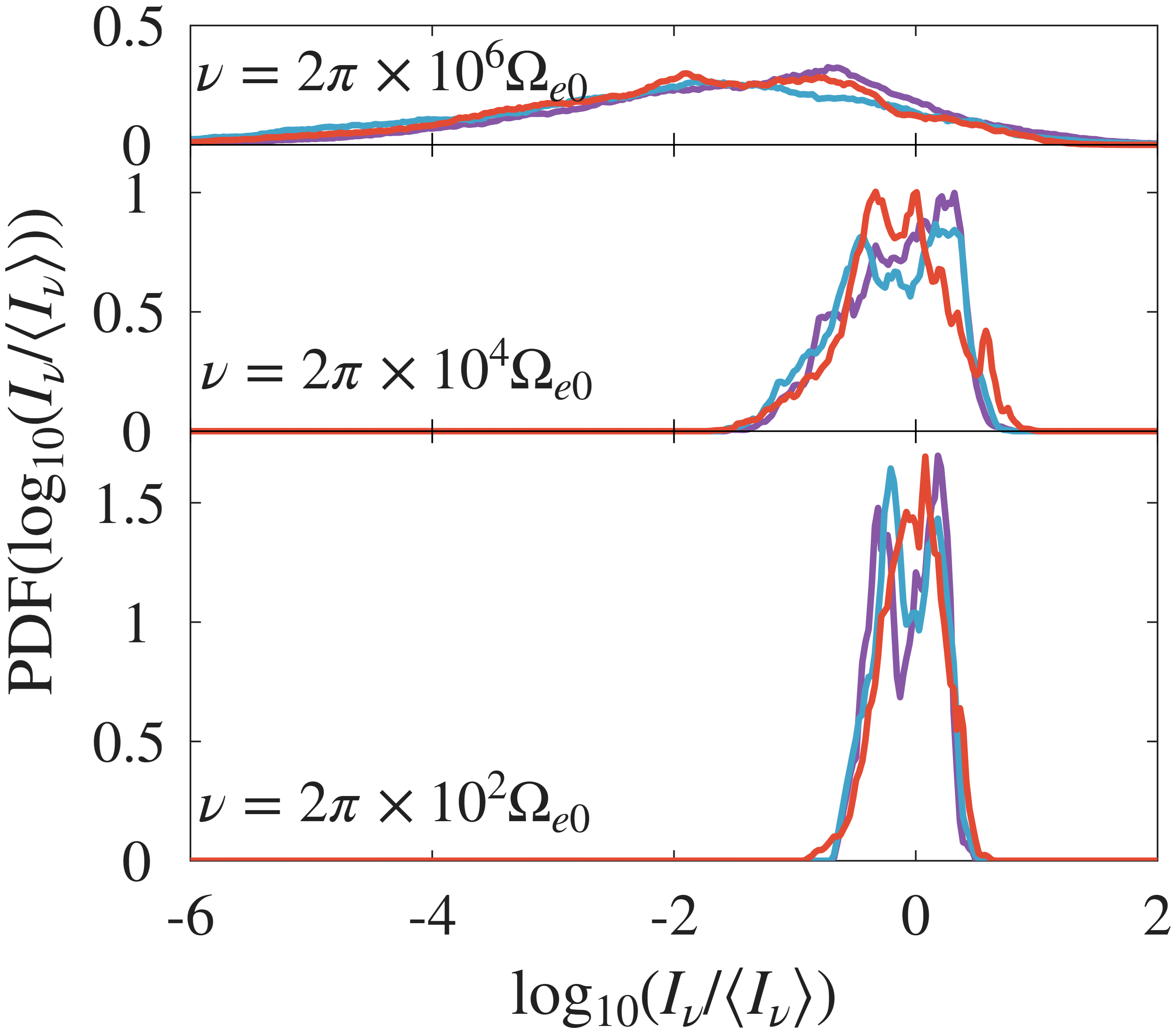}
    \caption{Probability distribution functions (PDFs) of synchrotron intensity variations $I_\nu / \langle I_\nu \rangle$ from quasi-steady turbulence with $\eta_\text{syn} = 6.4 \times 10^{-3}$, viewed along the $x$ direction, normalized by the spatial and temporal average $\langle I_\nu \rangle$ (Figure~\ref{fig::emission_total}). Colors denote different time snapshots, $t c / L = 6.25$ (purple), $7.8125$ (blue), and $9.375$ (red), while curves from bottom to top correspond to increasing observing frequency.}
    \label{fig::emission_pdf}
\end{figure}
\begin{figure*} 
    \centering
    \includegraphics[width=\textwidth]{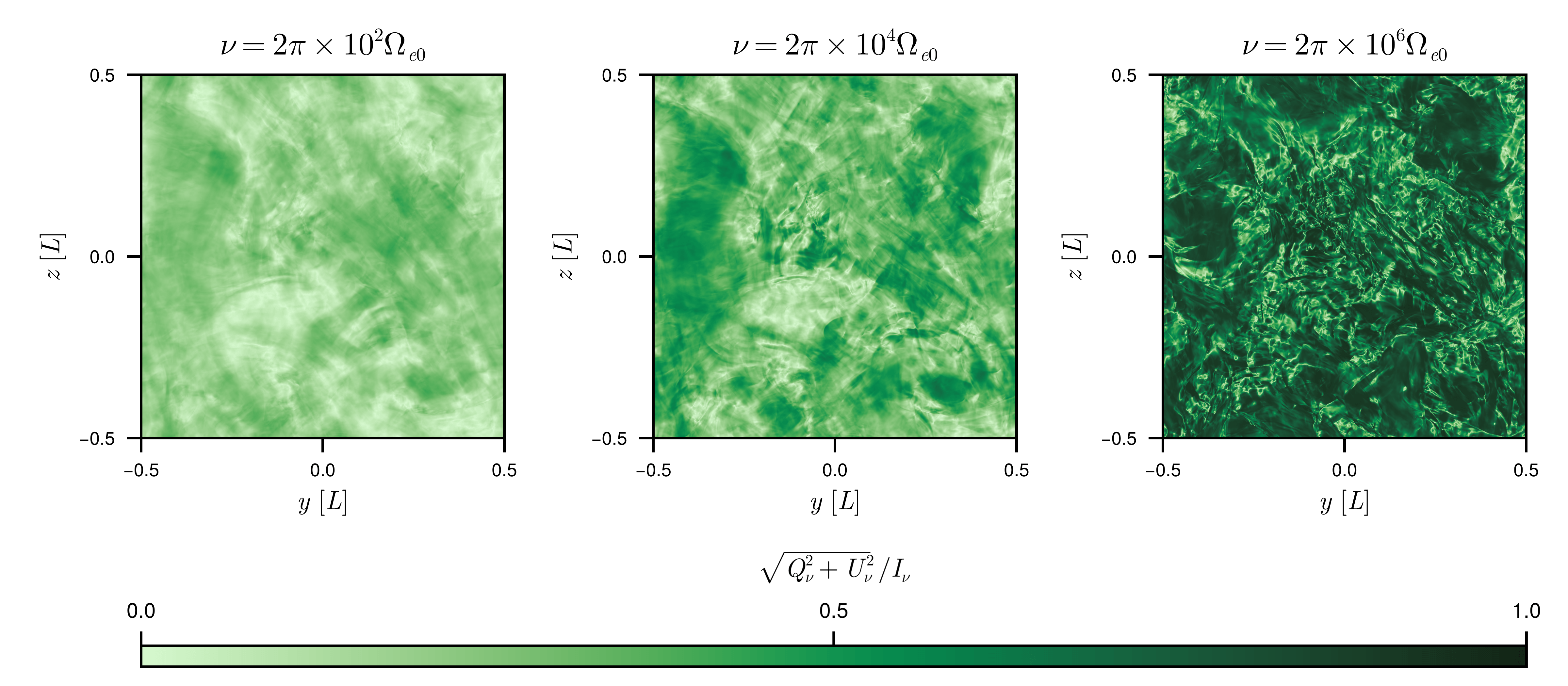}
    \caption{The linear polarization degree $\sqrt{Q_\nu^2 + U_\nu^2} / I_\nu$ images of turbulence with $\eta_\text{syn} = 6.4 \times 10^{-3}$, viewed along the $x$ direction at $t c / L = 9.375$. Panels from left to right show emission from low to high frequencies, as marked on the plot.}
    \label{fig::syn_pol_frac}
\end{figure*}
\begin{figure}
    \centering
    \includegraphics[width=0.45\textwidth]{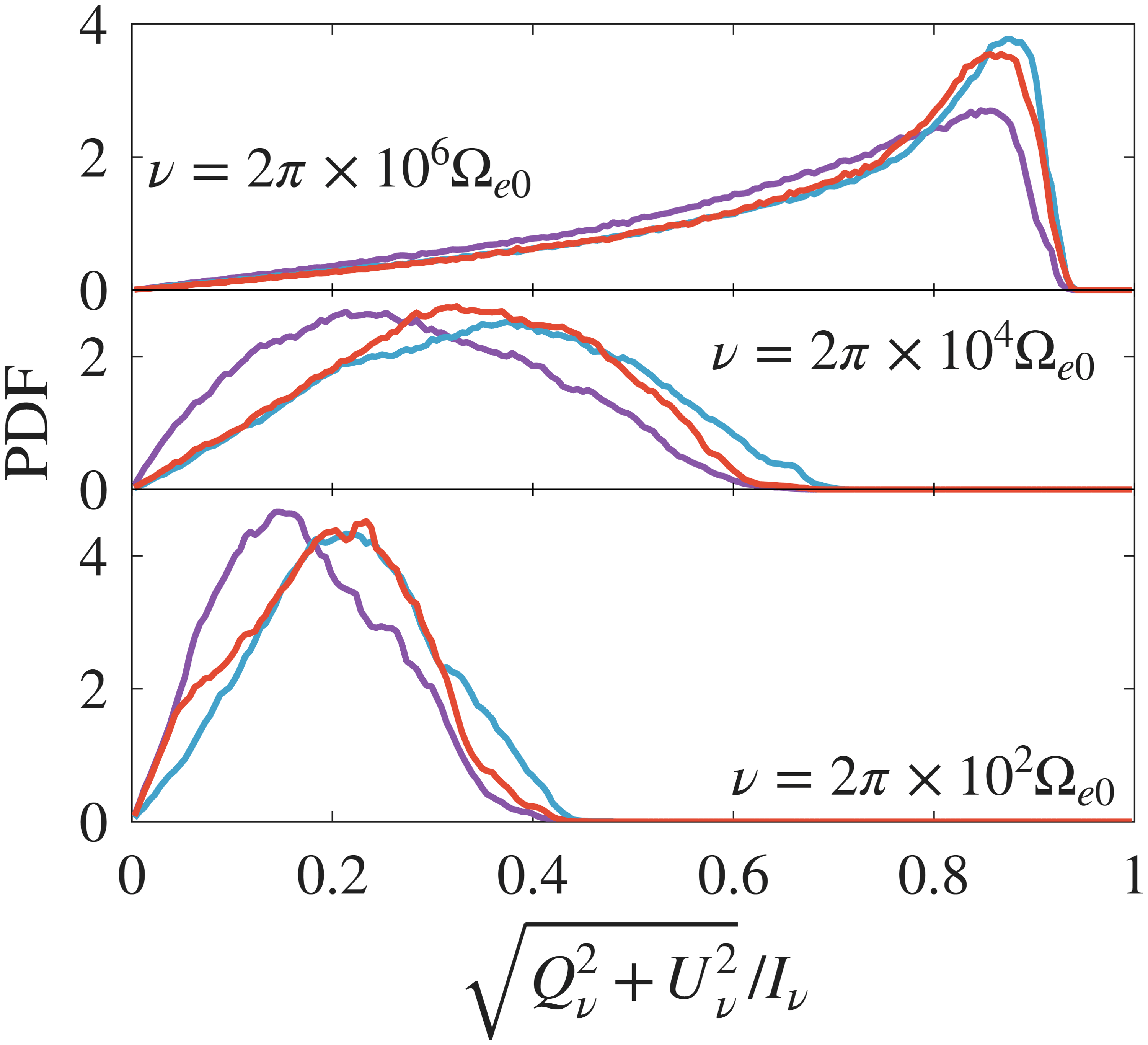}
    \caption{Probability distribution functions (PDFs) of synchrotron linear polarization degree $\sqrt{Q_\nu^2 + U_\nu^2} / I_\nu$ from quasi-steady turbulence with $\eta_\text{syn} = 6.4 \times 10^{-3}$, viewed along the $x$ direction. Colors denote different time snapshots, $t c / L = 6.25$ (purple), $7.8125$ (blue), and $9.375$ (red), while curves from bottom to top correspond to increasing observing frequency.}
    \label{fig::pol_frac_pdf}
\end{figure}
As shown in the right column of Figure~\ref{fig::snapshot}, the synchrotron cooling power $\gamma P_\text{syn}$ is spatially non-uniform. These spatial variations are most pronounced in the highly radiation-efficient case. Due to the inherent frequency dependence of synchrotron emission (Figure~\ref{fig::emission_total}), the effect of these spatial fluctuations depends on the observing frequency. To illustrate this, Figures~\ref{fig::syn_image} and \ref{fig::syn_pol_frac} visualize a representative case ($\eta_\text{syn}=6.4\times10^{-3}$) viewed along the $x$-axis, displaying the normalized synchrotron intensity $I_\nu/\langle I_\nu\rangle$ and the linear polarization degree $\sqrt{Q_\nu^2+U_\nu^2}/I_\nu$, respectively. Since the emission characteristics remain qualitatively similar for different viewing angles (Figure~\ref{fig::emission_total}), the following analysis mainly focuses on the spatial and temporal variability, as well as their frequency dependence, after fixing the LoS direction perpendicular to the mean magnetic field.

We compute the probability distribution function (PDF) of the synchrotron intensity variation $I_\nu / \langle I_\nu \rangle$ for three snapshots at different times (Figure~\ref{fig::emission_pdf}). At low frequencies, the emission originates from the entire plasma, so $I_\nu$ is always close to the mean value $\langle I_\nu \rangle$ (different values of $I_\nu$ are computed for different LoS cutting through the $y$-$z$ plane). With increasing frequency, the emission becomes dominated by rare, hot plasma columns with strong magnetic fields, producing broader non-Gaussian PDFs. At very high frequencies (e.g., $\nu = 2\pi \times 10^{6}\Omega_{e0}$), these hot, magnetized structures become directly visible (Figure~\ref{fig::syn_image}), whereas at lower frequencies they are masked by emission from cooler plasma. Their emission is highly intermittent, and the resulting PDFs vary significantly between different time snapshots (Figure~\ref{fig::emission_pdf}).

The variability of the linear polarization degree $\sqrt{Q_\nu^2+U_\nu^2}/I_\nu$ also exhibits a systematic frequency dependence. At low frequencies, the polarization degree (integrated along a given LoS cutting through the $y$-$z$ plane) is small for all choices of the LoS (Figure~\ref{fig::syn_pol_frac}). With increasing frequency, the emission becomes dominated by fewer plasma columns, suppressing depolarization and enhancing temporal variability. As a result, the polarization degree increases and displays intermittent, non-Gaussian PDFs (Figure~\ref{fig::pol_frac_pdf}). At high frequencies, individual plasma columns dominate most LoS, allowing the polarization degree to approach its theoretical maximum of $\sim72\%$ for our assumed electron distribution function (Equation~\ref{equ::emissivity_fit}). 

\subsection{Propagation effects: Faraday rotation}
\begin{figure*}
    \centering
    \includegraphics[width=\textwidth]{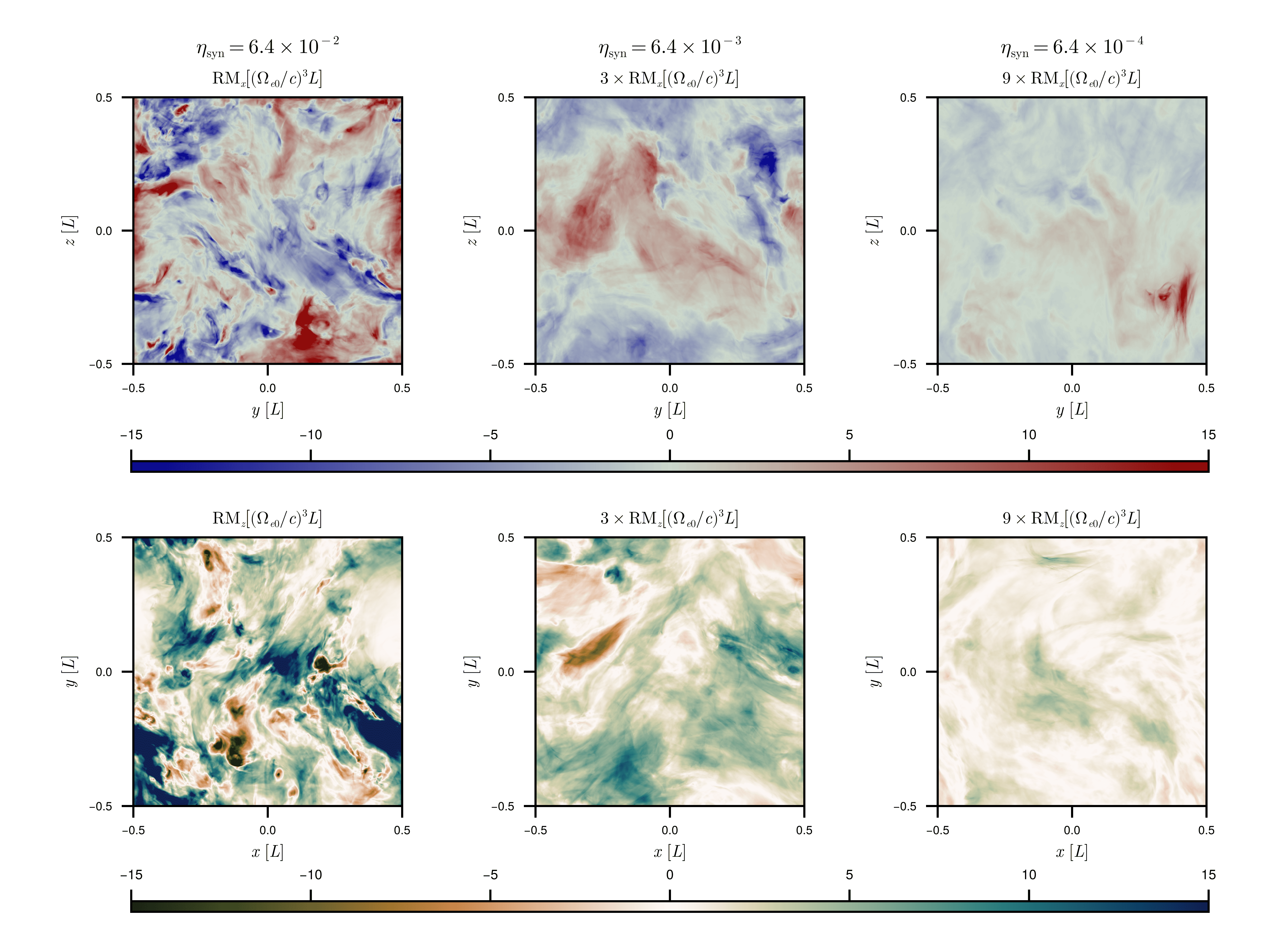}
    \caption{Rotation measure (RM) snapshots for turbulence with different cooling efficiencies. The top row shows views perpendicular to the mean magnetic field, and the bottom row shows views along the mean field. Note that, for better visualization, the RM in the $\eta_\text{syn} = 6.4 \times 10^{-3}$ case is multiplied by $3$, and the RM in the $\eta_\text{syn} = 6.4 \times 10^{-4}$ case is multiplied by $9$. All snapshots correspond to the same time $t c / L = 9.375$. }
    \label{fig::rm_map}
\end{figure*}
\begin{figure}
    \centering
    \includegraphics[width=0.45\textwidth]{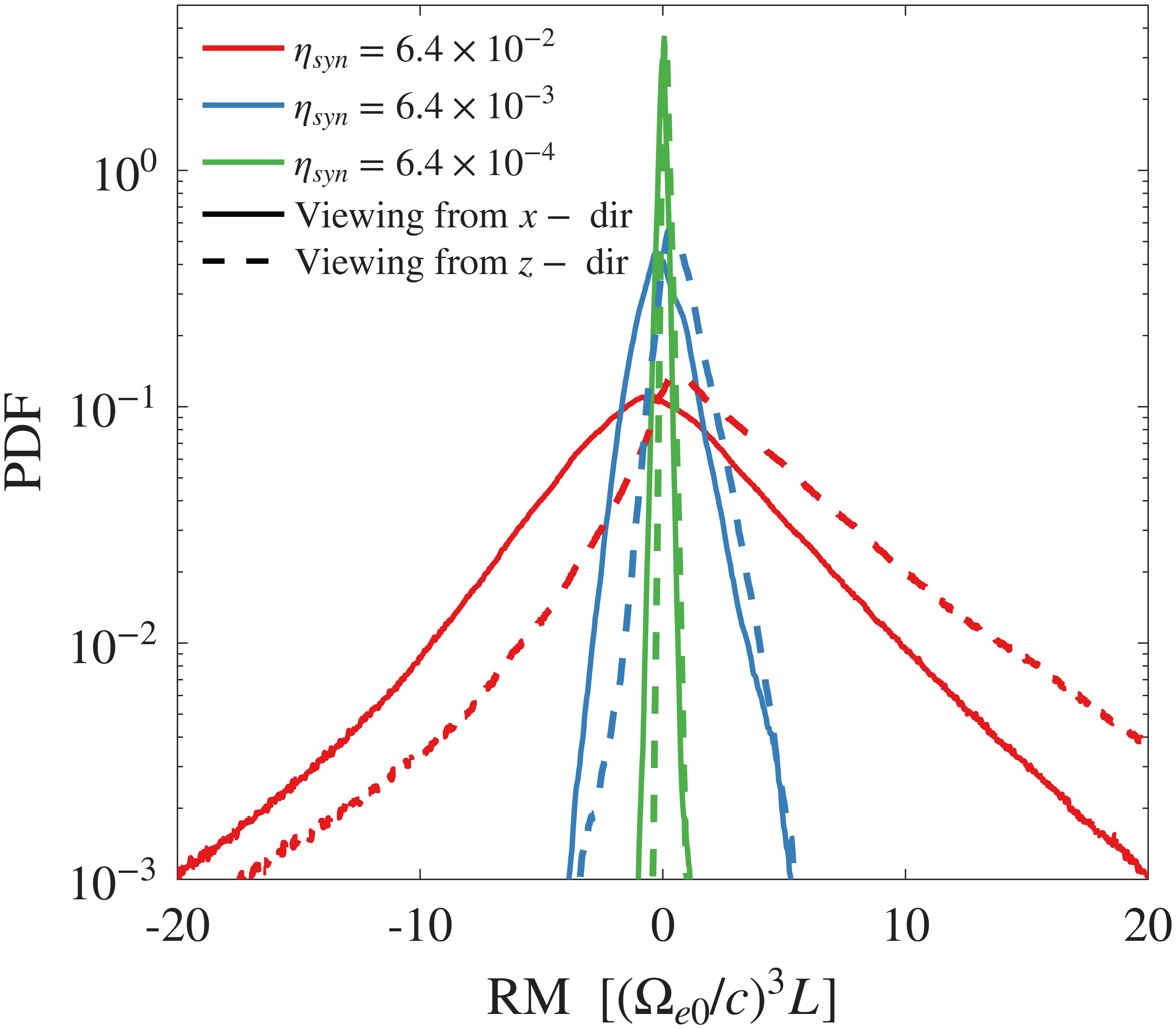}
    \caption{Probability distribution functions (PDFs) of the rotation measure (RM) from quasi-steady turbulence. Colors indicate different cooling efficiencies, as indicated in the legend. Solid (dashed) curves correspond to viewing directions perpendicular (parallel) to the mean magnetic field. The RM statistics is accumulated over $t c / L = 6.25$-$9.375$.}
    \label{fig::rm_pdf}
\end{figure}
When a linearly polarized wave propagates through a proton-electron plasma, its linear polarization angle rotates due to Faraday rotation \citep{1984frh..book.....M}:
\begin{align}
    \frac{\dd Q_\nu}{\dd s} = - U_\nu \times  \rho'_V \left(\frac{\nu}{\mathcal{D}}\right) / \mathcal{D} \, , \notag \\
    \frac{\dd U_\nu}{\dd s} =  Q_\nu \times \rho'_V \left(\frac{\nu}{\mathcal{D}}\right) / \mathcal{D} \, . \label{equ::faraday_rotation}
\end{align}
Here $\rho'_V$ is the Faraday rotativity in the fluid co-moving frame. We neglect Faraday conversion ($\rho'_Q \ll \rho'_V$ when $\nu' \gg \Omega_{e} $), synchrotron absorption, and light-travel-time delays within a single simulation snapshot. Lorentz contraction leads to $\rho'_V \rightarrow \rho'_V/\mathcal{D}$. For an linear polarized wave whose frequency in the fluid co-moving frame $\nu' \gg \Omega_{e}$, the Faraday rotation coefficient for a Maxwell-Jüttner distribution scales inversely with the square of the frequency \citep{2008ApJ...688..695S,2011MNRAS.410.1052S,2011MNRAS.416.2574H}, implying a polarization rotation angle proportional to wavelength squared. The proportionality constant defines the rotation measure (RM):
\begin{align}
    \text{RM} = 2 \pi^2  \left(\frac{\Omega_{e0}}{c} \right)^3 L \int_\text{LoS} \frac{\dd s}{L}  \frac{\rho}{\rho_0} \frac{B' \cos \theta}{\sqrt{\rho_0 c^2}} \frac{K_0\left(\Theta_t^{-1}\right)}{K_2\left(\Theta_t^{-1}\right)} \mathcal{D} \, , \label{equ::rm}
\end{align}
where the modified Bessel functions $K_0$ and $K_2$ incorporate relativistic corrections associated with the electron distribution. Here, we assume idealized observational conditions with arbitrarily high spatial and temporal resolution \citep{2022MNRAS.510.4654B}. We consider two different observer directions as in Section~\ref{sec::emission}. For each viewing direction, we integrate along the line of sight over the entire box.

Figure~\ref{fig::rm_map} shows RM snapshots from the three turbulence runs. When viewed from either the $x$- or the $z$-direction, the RM fluctuates. Its variance depends strongly on the cooling efficiency: the cooling-efficient turbulence ($\eta_\text{syn}=6.4\times10^{-2}$) tends to produce larger RM fluctuations, whereas the cooling-inefficient case ($\eta_\text{syn}=6.4\times10^{-4}$) produces much smaller RM values.

Figure~\ref{fig::rm_pdf} summarizes the spatiotemporal statistics of the RM and further demonstrates its strong dependence on the cooling efficiency. In all runs, the RM distributions approach Laplace distributions, with exponentially decaying tails at large RM values \citep[see also high-Mach-number turbulence in ][]{2021MNRAS.502.2220S}. We quantify the Faraday rotation by the standard deviation (S.D.) of the RM, listed in Table~\ref{tab::rm_variance}, to compare different turbulence runs. The S.D. of the RM is systematically larger in cooling-efficient turbulence than in cooling-inefficient cases.
The dependence on cooling efficiency arises primarily because cooling-inefficient turbulence leads to hotter plasma (Figure~\ref{fig::phase}), which suppresses Faraday rotation \citep[e.g.,][]{1979ApJ...228..268J, 2000ApJ...545..842Q, 2008ApJ...688..695S}. Based on Table~\ref{tab::rm_variance}, the S.D. of the RM broadly scales with volume-averaged quantities as
\begin{align}
    \text{S.D.}\left(\text{RM}\right) \propto  \sqrt{\left\langle \sigma \right\rangle_V}   \frac{K_0\left(\left\langle \Theta_t \right\rangle_V^{-1}\right)}{K_2\left(\left\langle \Theta_t \right\rangle_V^{-1}\right)}. \label{equ::rm_scaling}
\end{align}
Since the mean density and mean magnetic field strength are similar among our runs (Figure~\ref{fig::time}), we conclude that the RM dependence is mainly controlled by the mean temperature (Table~\ref{tab::rm_variance}). The importance of correlations between density, magnetic field, and temperature fluctuations, induced by the thermal instability (\citealp{1992A&A...256..689B};  see also Appendix~\ref{sec::thermal_ins}) appears subdominant compared to the role of the mean temperature \citep{2009ApJ...705L..86W, 2021MNRAS.502.2220S}, possibly because the RM is an integrated measure over several uncorrelated turbulent eddies (Equation~\ref{equ::rm}).

The rotation measure $\text{RM}_z$ viewed along the mean magnetic field direction ($z$-direction) is preferentially positive, whereas $\text{RM}_x$ remains approximately symmetric around zero. The mean-field contribution to the Faraday rotation, quantified by $\langle \text{RM}_z \rangle$ in Table~\ref{tab::rm_variance}, is comparable to the turbulent contribution, measured by S.D.$(\text{RM}_z)$ or S.D.$(\text{RM}_x)$. Thus, Faraday rotation remains sensitive to the mean-field direction, unlike synchrotron emission and its linear polarization which are largely independent of the LoS direction (Section~\ref{sec::emission}).

Faraday conversion between linear and circular polarization may provide an additional probe of turbulence properties \citep[e.g.,][]{2002ApJ...573..485R, 2019ApJ...876...74G}; however, we leave it for future work.
\begin{table}
    \centering
    \begin{tabular}{c|c|c|c}
        \hline
        $\eta_\text{syn}$ & $6.4 \times 10^{-2}$ & $6.4 \times 10^{-3}$ & $6.4 \times 10^{-4}$ \\
        \hline
        $\left\langle\text{RM}_z \right\rangle [(\Omega_{e0}/c)^3 L]$  & 2.7 & 0.73 & 0.13 \\
        S.D.$\left(\text{RM}_x\right) [(\Omega_{e0}/c)^3 L]$ & 5.2 & 1.1 & 0.17 \\
        S.D.$\left(\text{RM}_z\right) [(\Omega_{e0}/c)^3 L]$ & 5.4 & 0.97 & 0.13 \\
        $\frac{K_0\left(\langle \Theta_t \rangle_V^{-1}\right)}{K_2\left(\langle \Theta_t \rangle_V^{-1}\right)}$ & 1.2e-1 & 2.5e-2 & 5.8e-3 \\
        \hline
    \end{tabular}
    \caption{The mean ($\langle \rangle$) and standard deviation (S.D.) of the rotation measure (RM) for different cooling efficiencies $\eta_\text{syn}$ in quasi-steady turbulence, from time $6$ to $9.735\,L/c$.}
    \label{tab::rm_variance}
\end{table}

\section{Comparisons with previous relativistic magnetized turbulence simulations}
\label{sec::compare}
\citet{2012ApJ...744...32Z} performed the first relativistic MHD simulations of driven turbulence using an external acceleration applied to the velocity field, similar to our stirring force $\boldsymbol{F}$ (Equation~\ref{equ::net_force}). We adopt the same scheme because it explicitly conserves the total momentum and yields an approximately constant energy injection rate, $F_t \simeq \boldsymbol{F} \cdot \boldsymbol{M} / (w_\text{lab}c)$ (Appendix~\ref{sec::turb_driving}). Unlike \citet{2012ApJ...744...32Z}, our simulations include radiative cooling.

We note that the imposed perturbations are prescribed in Fourier space and applied simultaneously, so causality is not strictly enforced; this limitation is shared by most relativistic turbulence simulations \citep{2012ApJ...744...32Z, 2013ApJ...763L..12Z, 2021ApJ...922..172Z, 2023PhRvL.131e5201S, 2025arXiv250303820G}. We also neglect explicit resistivity, although reconnection may influence the dissipation range and the development of the thermal instability \citep{2018PhRvL.121y5101C,2019ApJ...886..122C,2021PhRvL.127y5102C,2021ApJ...923L..13C,2025ARA&A..63..127S}. 

The inertial-range cascade is largely insensitive to the specific choices of driving and dissipation prescriptions (Figure~\ref{fig::pwr}): previous decaying relativistic MHD and force-free simulations also find Kolmogorov-type magnetic spectra in the inertial range \citep{2011ApJ...734...77I, 2016ApJ...831L..11T, 2021ApJ...923L..13C, 2005ApJ...621..324C, 2016ApJ...817...89Z}.

\section{Implications for astrophysical sources}
\label{sec::application}
Turbulence accompanied by strong synchrotron cooling is expected in many high-energy astrophysical systems, including AGN jets \citep[e.g.,][]{2019ARA&A..57..467B}, accretion flows \citep[e.g.,][]{2014ARA&A..52..529Y}, pulsar wind nebulae \citep[e.g.,][]{2017SSRv..207..137P}, and fast radio bursts \citep[e.g.,][]{2023RvMP...95c5005Z}. In this subsection, we discuss the observational implications of our results for these environments. 
\subsection{Stochastic synchrotron flares in blazar jets and accretion flows}
Blazars exhibit rapid synchrotron variability on timescales from minutes to years \citep[e.g.,][]{2025A&ARv..33....8R}. Recent multi-wavelength observations of Mrk~421 and Mrk~501 show synchronous optical and X-ray flares, with greater flux variability and stronger polarization enhancements at higher frequencies \citep{2022Natur.611..677L, 2025ApJ...986..230M, 2025A&A...695A.217M, 2025A&A...703A..19C}. Such stochastic flux and polarization variability is commonly attributed to magnetized turbulence in the blazar jet \citep[e.g.,][]{2014ApJ...780...87M, 2017Galax...5...63M} and qualitatively resembles the results of our simulations (Figures~\ref{fig::emission_pdf}~\&~\ref{fig::pol_frac_pdf}), where the low-frequency polarization degree stays small while the high-frequency polarization degree intermittently approaches the theoretical maximum. Future multi-wavelength polarimetry may reveal similar behavior in other AGN jets \citep[e.g.,][]{2015A&A...578A..68C, 2023MNRAS.523.3615Z}.

Infrared variability in Sgr~A* is often interpreted as non-thermal synchrotron emission from transient hot spots in the inner accretion flow \citep{2006ApJ...640L.163G, 2014IAUS..303..274W, 2018ApJ...863...15W, 2025ApJ...979L..20V, 2025arXiv251114836M}. Such stochastic variability may arise from relativistic turbulence \citep[e.g. ][]{2023PhRvR...5d3023Z, 2025PhRvD.112l3028L}, possibly in synergy with the thermal instability, which can intermittently produce hot spots on timescales comparable to (or shorter than) the eddy turnover time.

\subsection{X-ray polarization in pulsar wind nebulae}
At the termination shock, the ordered motion of pulsar winds transitions into a chaotic, turbulent, magnetized flow within the nebula
\citep[e.g.,][]{2014MNRAS.438..278P, 2014RPPh...77f6901B, 2016JPlPh..82d6301L}. In the inner nebula, turbulence dissipates magnetic energy and accelerates pairs that radiate synchrotron X-rays \citep[e.g.,][]{2016ApJ...823...39Z}. Recent X-ray polarimetry of the Crab and Vela nebulae shows a spatially inhomogeneous polarization degree that increases with photon energy \citep{2022Natur.612..658X, 2023NatAs...7..602B,2023ApJ...959L...2L, 2024ApJ...968L..18T}, qualitatively consistent with our results (Figures~\ref{fig::syn_pol_frac}~\&~\ref{fig::pol_frac_pdf}): upon integration along the line of sight, low-frequency emission is depolarized by disordered magnetic fields; instead, high-frequency emission is dominated by rare, hot, strongly magnetized columns and therefore reaches higher polarization degrees \citep[see also][]{2025A&A...702A.171D}.

However, the observed polarization degree ($\gtrsim 40\%$) and  spectral modeling of the Crab and Vela nebulae suggest that the turbulent field is comparable to the mean field \citep[e.g.,][]{2017MNRAS.470.4066B, 2019A&A...627A.100H}. Our simulations currently probe a stronger-turbulence regime (turbulent magnetic energy $\gtrsim 10\times$ the initial mean field energy; Figure~\ref{fig::time}); extension to the weak-turbulence limit is deferred to future work.

\subsection{Faraday rotation in Fast Radio Bursts}
The rotation measures of several repeating FRBs display significant temporal variability, on timescales of days, plausibly linked to a turbulent, magnetized  nebula surrounding the FRB source \citep[e.g.,][]{2018Natur.553..182M, 2022NatCo..13.4382W, 2023Sci...380..599A, 2025A&A...694A..75B}. Our results in Figures~\ref{fig::rm_map} and \ref{fig::rm_pdf} likewise show that the RM can fluctuate strongly  in turbulence. A more quantitative comparison with observations, e.g., through the power-law slopes of structure functions and Fourier spectra of the two-dimensional RM maps (Figure~\ref{fig::rm_map}), is deferred to future work \citep[e.g.,][]{2021ApJ...910...88X, 2023MNRAS.518..919S, 2025ApJ...979L..41L}.

Our RM scaling relation (Equation~\ref{equ::rm_scaling}) quantitatively constrains the Faraday screen, especially its temperature. Consistent with previous theoretical estimates for FRB 121102 \citep[e.g.,][]{2018ApJ...868L...4M},
which has an observed RM of $\sim 1.46 \times 10^5 \text{rad}\ \text{m}^{-2}$, if we take the electron number density $\sim 100\, \text{cm}^{-3}$, the magnetization $\langle \sigma \rangle_V  \sim 0.1$ and the size of the Faraday screen $\sim10^{17} \text{cm}$, the average electron temperature would be $\langle \Theta_t \rangle_V  \sim 3 \times 10^3$.

\section{Conclusions}
\label{sec::conclusion}
We perform the first relativistic MHD simulations of driven turbulence with synchrotron cooling. Energy is injected at the largest scale, cascades to smaller scales, and is finally radiated away by synchrotron emission after numerical dissipation into heat. Through the combined action of turbulence and synchrotron cooling, the plasma thermodynamics is self-consistently regulated, and the total cooling power must balance the energy injection rate in quasi-steady state (Figure~\ref{fig::time}).

By varying the synchrotron cooling efficiency $\eta_\text{syn}$ (Equation~\ref{equ::cooling}) while fixing the driving, we find:
\begin{itemize}
    \item The volume-averaged plasma temperature $\left\langle \Theta_t \right \rangle_V$ drops as $\eta_\text{syn}$ increases, as $\left\langle \Theta_t \right \rangle_V \propto \eta_\text{syn}^{-1/2}$ (Figure~\ref{fig::time}),  in agreement with theoretical expectations \citep{2018MNRAS.477.2849U}. 
    \item The thermal instability induced by synchrotron cooling \citep{1967ApJ...150..105S,1979ApJ...233..463E} bifurcates the plasma into a cold dense phase and a hot dilute phase while competing with turbulent mixing (Figure~\ref{fig::phase}~\&~\ref{fig::phase_identification_snapshot}).
    \item The magnetic power spectral density (Figure~\ref{fig::pwr})  follows a Kolmogorov-like scaling ($\sim k^{-5/3}$) within the inertial range, consistent with previous studies of relativistic MHD/force-free turbulence without synchrotron cooling \citep[e.g.,][]{2012ApJ...744...32Z, 2021ApJ...923L..13C}.
    \item The integrated synchrotron spectrum is rather broad, as the thermal plasma spans a wide range of temperatures (Figure~\ref{fig::emission_total}). Tangled magnetic fields depolarize the low-frequency emission integrated along the LoS. In contrast, the high-frequency emission, dominated by rare, hot, strongly-magnetized columns, approaches the maximum linear polarization degree. This is consistent with the trend seen in the Vela nebula \citep{2022Natur.612..658X}.
    \item The total intensity and the linear polarization degree display stronger spatial and temporal variability at higher frequencies (Figures~\ref{fig::emission_pdf}~\&~\ref{fig::pol_frac_pdf}), qualitatively resembling the flaring behavior of blazars and Sgr~A*
    , where synchrotron flux variability increases with observing frequency.
\citep{2022Natur.611..677L,2025ApJ...986..230M,2025A&A...695A.217M,2025A&A...703A..19C}.
    \item The rotation measure (RM) fluctuates in space and time (Figure~\ref{fig::rm_pdf}). The mean-field contribution to the
Faraday rotation is
comparable to the turbulent contribution, and they both decrease for higher plasma temperatures (Table~\ref{tab::rm_variance}). Our calculations support previous estimates that a highly magnetized relativistically hot ion-electron plasma can serve as the Faraday screen around FRB sources \citep{2018ApJ...868L...4M}.
\end{itemize}

We acknowledge several limitations, including our choice of a fixed driving scheme (Appendix~\ref{sec::turb_driving}) and the lack of explicit resistivity (Section~\ref{sec::compare}) or thermal conduction. A more realistic radiation model should include nonthermal electrons \citep[e.g.,][]{2019ApJ...886..122C, 2019PhRvL.122e5101Z, 2021ApJ...921...87N, 2023ApJ...949...71Z}, anisotropic electron distributions \citep[e.g.,][]{2020ApJ...895L..40C, 2021PhRvL.127y5102C, 2022ApJ...936L..27C, 2023ApJ...957..103G}, a two-temperature plasma \citep[e.g.,][]{1995ApJ...452..710N, 2016A&A...586A..38M, 2023ApJ...944...24Z}, anisotropic synchrotron emissivity, and additional radiative processes such as synchrotron self-absorption and inverse Compton scattering \citep[e.g.,][]{2020MNRAS.493..603Z, 2021ApJ...921...87N, 2021MNRAS.503..688S, 2024NatCo..15.7026N, 2024PhRvL.132h5202G, 2025ApJ...987..159M}. Future studies will address these limitations and incorporate the evolution of nonthermal particles in quasi-steady relativistic turbulence.

\begin{acknowledgments} 
We acknowledge fruitful discussions with Xue-Ning Bai, James R. Beattie, Amitava Bhattacharjee, Sasha Chernoglazov, Jordy Davelaar, Chang-Goo Kim, Brian Metzger and Jonathan Zrake. This work was supported by a grant from the Simons Foundation (MP-SCMPS-00001470) and facilitated by Multimessenger Plasma Physics Center (MPPC), NSF grants PHY-2206607 and PHY-2206609. X.~S.~acknowledges support from NASA ATP grant 80NSSC22K0667 and NSF Grant No. PHY-2308944.  L.~C.~acknowledges support from NSF Grant No. PHY-2308944, NASA ATP Award No. 80NSSC22K0667, and NASA ATP Award No. 80NSSC24K1230. L.~S.~acknowledges support from DoE Early Career Award DE-SC0023015, NASA ATP 80NSSC24K1238, NASA ATP 80NSSC24K1826, and NSF AST-2307202. A.~P.~ acknowledges support by an Alfred P. Sloan Fellowship, and a Packard Foundation Fellowship in Science and Engineering. Simulations were performed on computational resources provided by the Princeton Research Computing, by the Flatiron Institute and by the National Energy Research Scientific Computing Center (NERSC), a Department of Energy User Facility using NERSC award DDR-ERCAP0034731.
\end{acknowledgments}

\appendix
\section{Numerical scheme for the turbulence driving force}
\label{sec::turb_driving}
This section describes the stirring force $\mathbb{F}$ used in our simulations and presents a benchmark test of its numerical implementation. The force is designed to inject momentum into gas motions and Poynting flux while preserving the plasma temperature $\Theta_t$. In the explicit numerical scheme, $\mathbb{F}$ modifies the conserved variables $\left(\boldsymbol{M}, \varepsilon\right)$.
After applying the stirring force, the momentum density updates as $\boldsymbol{M}_\text{new} = \boldsymbol{M}_\text{old} + \dd t\, \boldsymbol{F}$, where $\boldsymbol{F}$ is a random force vector. To enforce no direct gas heating and conservation of the lab-frame density ($\rho_\text{lab} = \gamma \rho$), the time-like component $F_t$, which controls the increment of the kinetic and electric field energy densities, is fully constrained by $\boldsymbol{M}_\text{new}$, $\boldsymbol{B}$, $\rho_\text{lab}$, and $w_{g,\text{lab}}\equiv \gamma w_g $ \citep[Eq.~5.4 in][]{2014SJSC...36B.661N},
\begin{align}
    \varepsilon_\text{new} = & \varepsilon_\text{old} + \dd t F_t \notag \\
     = &  \left|\boldsymbol{B}\right|^2/2 - \frac{\Gamma - 1}{\Gamma \gamma_\text{new}} \left(w_{g, \text{lab}} - \rho_\text{lab} c^2\right) + \gamma_\text{new} w_{g, \text{lab}} \notag \\ & + \frac{\left|\boldsymbol{B}\right|^2 \left|\boldsymbol{M}_\text{new}\right|^2 - \left(\boldsymbol{M}_\text{new}\cdot \boldsymbol{B}\right)^2}{2 \left(\left|\boldsymbol{B}\right|^2 + \gamma_\text{new} w_{g, \text{lab}} \right)^2} c^2, \label{equ::energy_driving}
\end{align}
where the post-stirring Lorentz factor $\gamma_\text{new}$ satisfies a quartic equation \citep[Eq~4.8 in][]{2014SJSC...36B.661N},
\begin{align}
    &\gamma_\text{new}^4 +  \frac{2\left|\boldsymbol{B}\right|^2}{w_{g, \text{lab}}} \gamma_\text{new}^3 +\left[ \frac{\left|\boldsymbol{B}\right|^4}{w^2_\text{g,lab}} - \frac{\left|\boldsymbol{M}_\text{new}\right|^2 c^2}{w^2_\text{g,lab}} - 1 \right] \gamma_\text{new}^2 \notag \\
    &- \left[\frac{2\left|\boldsymbol{B}\right|^2}{w_{g, \text{lab}}} + \frac{2 \left(\boldsymbol{M}_\text{new}\cdot \boldsymbol{B}\right)^2 c^2}{w^3_\text{g,lab}}\right] \gamma_\text{new}  \notag  \\ = & \frac{\left|\boldsymbol{B}\right|^4}{w^2_\text{g,lab}} + \frac{\left|\boldsymbol{B}\right|^2 \left(\boldsymbol{M}_\text{new}\cdot \boldsymbol{B}\right)^2 c^2}{w^4_\text{g,lab}}. \label{equ::gamma_new}
\end{align}
We solve for $\gamma_\text{new} \geq 1$ using Schroeder's method.

\begin{figure}
    \centering
    \includegraphics[width=0.45\textwidth]{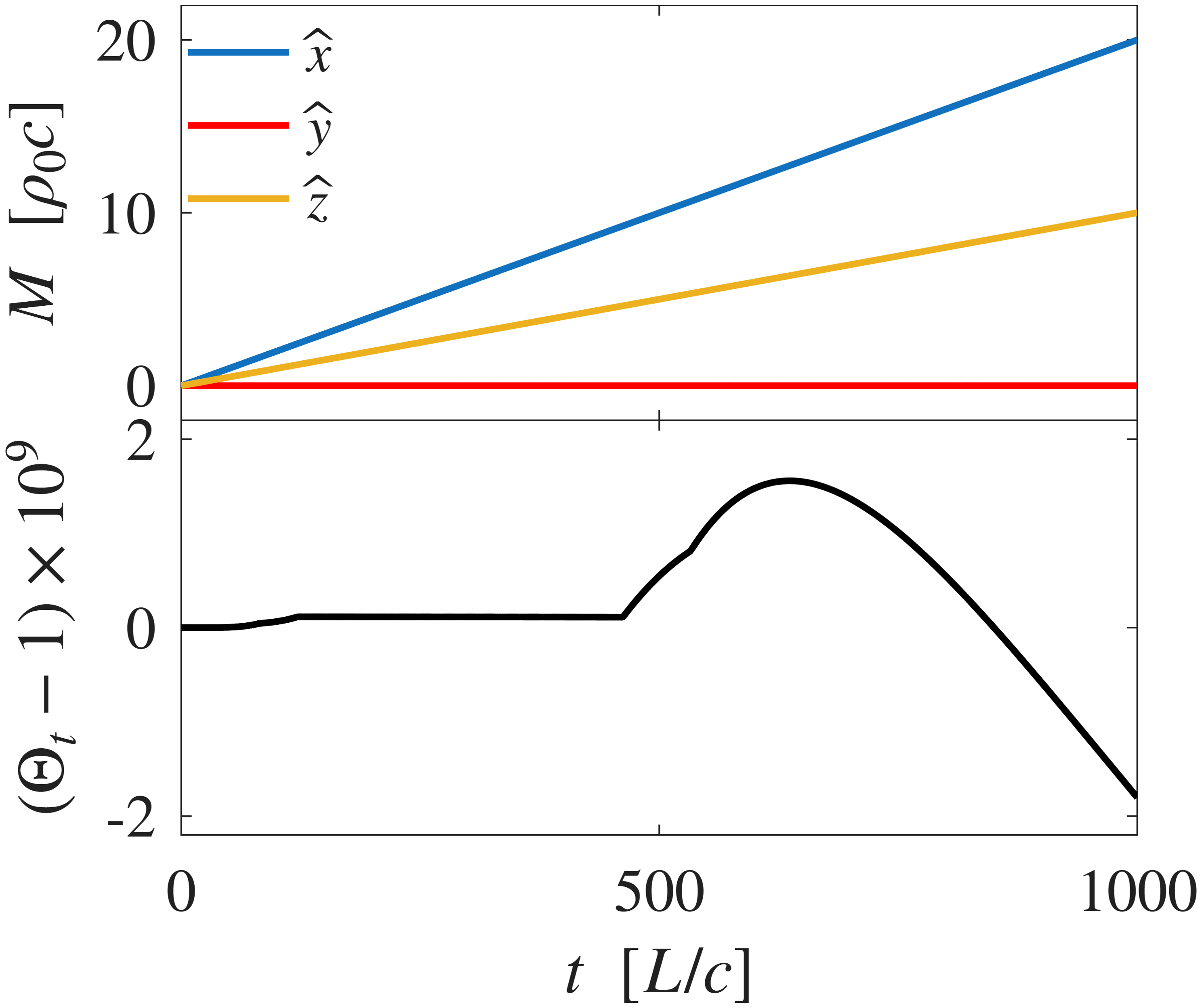}
    \caption{The benchmark test for the numerical implementation of stirring force $\mathbb{F}$ (Equation~\ref{equ::energy_driving}~\&~\ref{equ::gamma_new}). The top panel shows the time evolution of the total momentum density $\boldsymbol{M}$ where colors refer to different direction components. The bottom panel gives the time evolution of the gas temperature $\Theta_t$, which fluctuates around $1 \pm 2 \times 10^{-9}$.}
    \label{fig::benchmark}
\end{figure}

We perform a benchmark test to verify that there is no  variation in the internal energy of the gas (Equations~\ref{equ::energy_driving} and \ref{equ::gamma_new}). For this test, we consider a fully controlled setup for $\mathbb{F}$: a uniform static gas with initial density $\rho_0$, gas pressure $P_{g} = \rho_0 c^2$, carrying a uniform background magnetic field $B_0 \hat{z}$ where $B_0^2 = \rho_0 c^2$; a constant uniform stirring force $\boldsymbol{F} = (2 \hat{x} + \hat{z}) \times 10^{-2} \rho_0 c^2/L$, oblique to the magnetic field. Figure~\ref{fig::benchmark} shows the long-term evolution: the momentum density increases linearly up to $\left|\boldsymbol{M}\right| \sim \mathcal{O}\left(10\right)$, and the total energy density $\varepsilon$ increases accordingly. As expected, the temperature $\Theta_t$ remains unchanged within numerical precision ($\sim 2 \times 10^{-9}$), with minor deviations arising from numerical errors in solving the quartic equation (Equation~\ref{equ::gamma_new}) and computing $P_g$ from $\varepsilon$.

\citet{2012ApJ...744...32Z} employed an alternative driving scheme, by applying a random acceleration to the velocity field $\gamma\boldsymbol{u}$, which similarly injects momentum and Poynting flux into the flow. Our method directly controls the momentum density and explicitly conserves the total momentum, $\langle \boldsymbol{M} \rangle_V = 0$. We control the magnitude of the stirring force via $\langle \boldsymbol{M} \cdot \boldsymbol{F} \rangle_V$, rather than $\langle |\partial_t (\gamma\boldsymbol{u})| \rangle_V$ \citep{2012ApJ...744...32Z}. In the Newtonian limit, our choice yields an approximately constant energy injection rate, $\langle F_t \rangle_V \sim \langle \boldsymbol{M} \cdot \boldsymbol{F}_{OU} \rangle_V / \langle w_\text{lab} \rangle_V$.

\section{Thermal instability induced by synchrotron cooling: linear and non-linear evolution}
\label{sec::thermal_ins}
\begin{figure}
    \centering
    \includegraphics[width=0.45\textwidth]{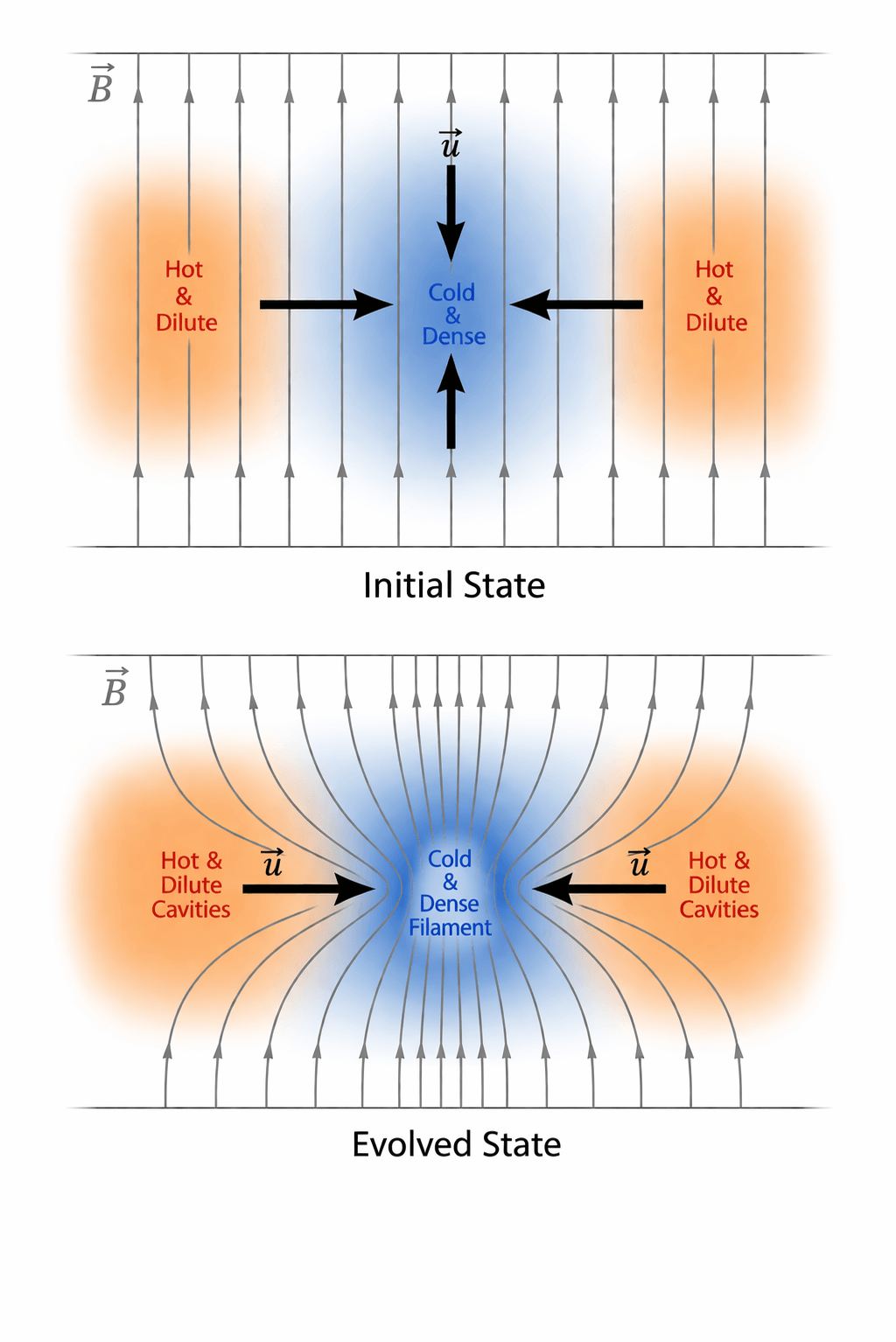}
    \caption{Illustration of the thermal instability induced by synchrotron cooling. This schematic illustration was generated using \texttt{ChatGPT (OpenAI)} based on author-provided physically-motivated prompts and refined by the authors.}
    \label{fig::cool_instability_carton}
\end{figure}
\begin{figure}
    \centering
    \includegraphics[width=0.45\textwidth]{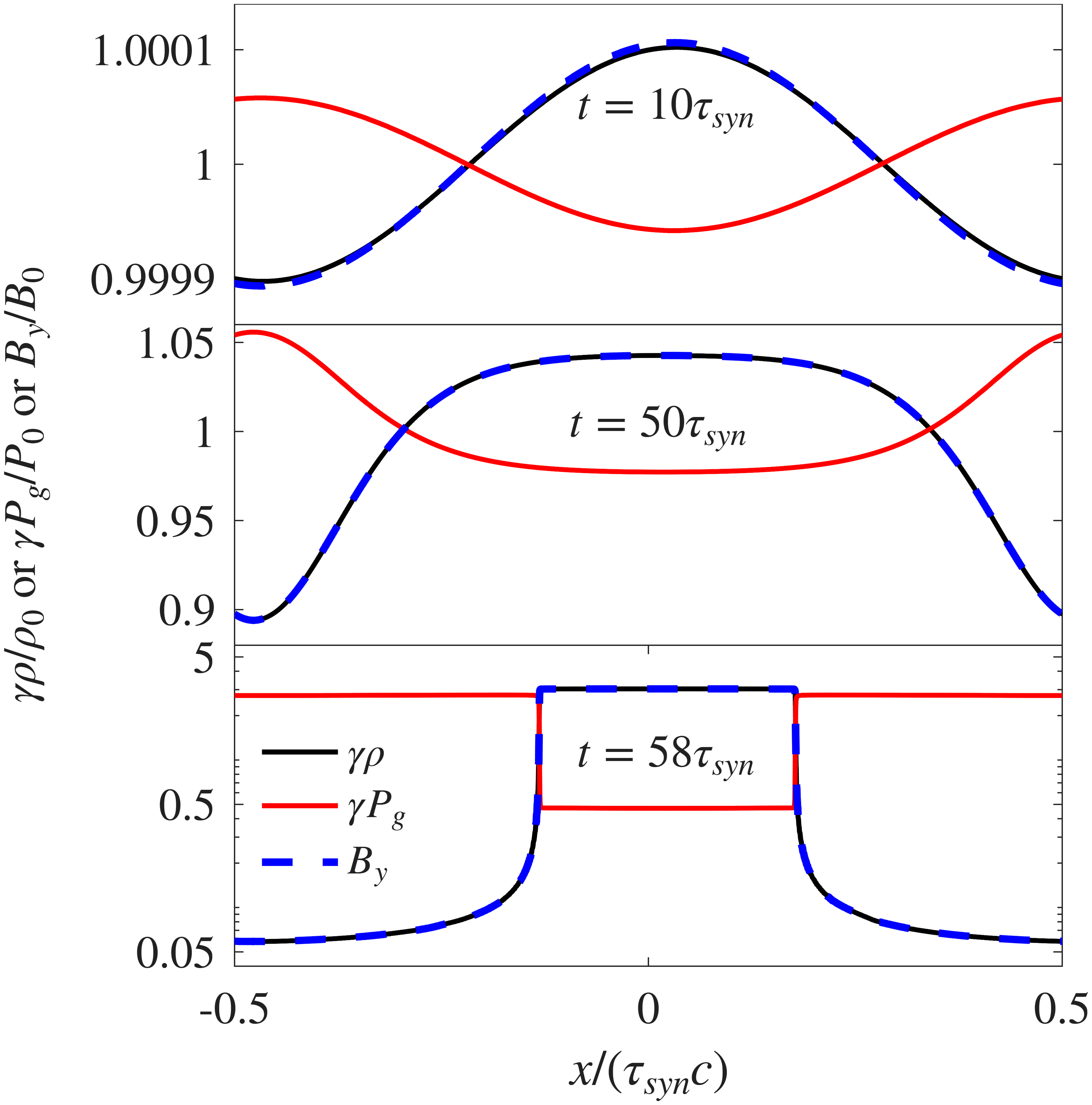}
    \caption{Single mode evolution of the thermal instability induced by synchrotron cooling, for lab frame density $\gamma \rho$ (black solid lines),  gas pressure $\gamma P_g$ (red solid lines), and magnetic field $B_y$ (blue dashed lines). The magnetic field curves largely overlap with the density profiles. Panels from top to bottom show the linear ($t = 10\, \tau_{\rm syn}$), quasi-linear ($t = 50\, \tau_{\rm syn}$), and non-linear ($t = 58\, \tau_{\rm syn}$) stages. Note that the vertical axis in the bottom panel is in logarithmic scale.}
    \label{fig::cool_instability_single}
\end{figure}

\begin{figure}
    \centering
    \includegraphics[width=0.45\textwidth]{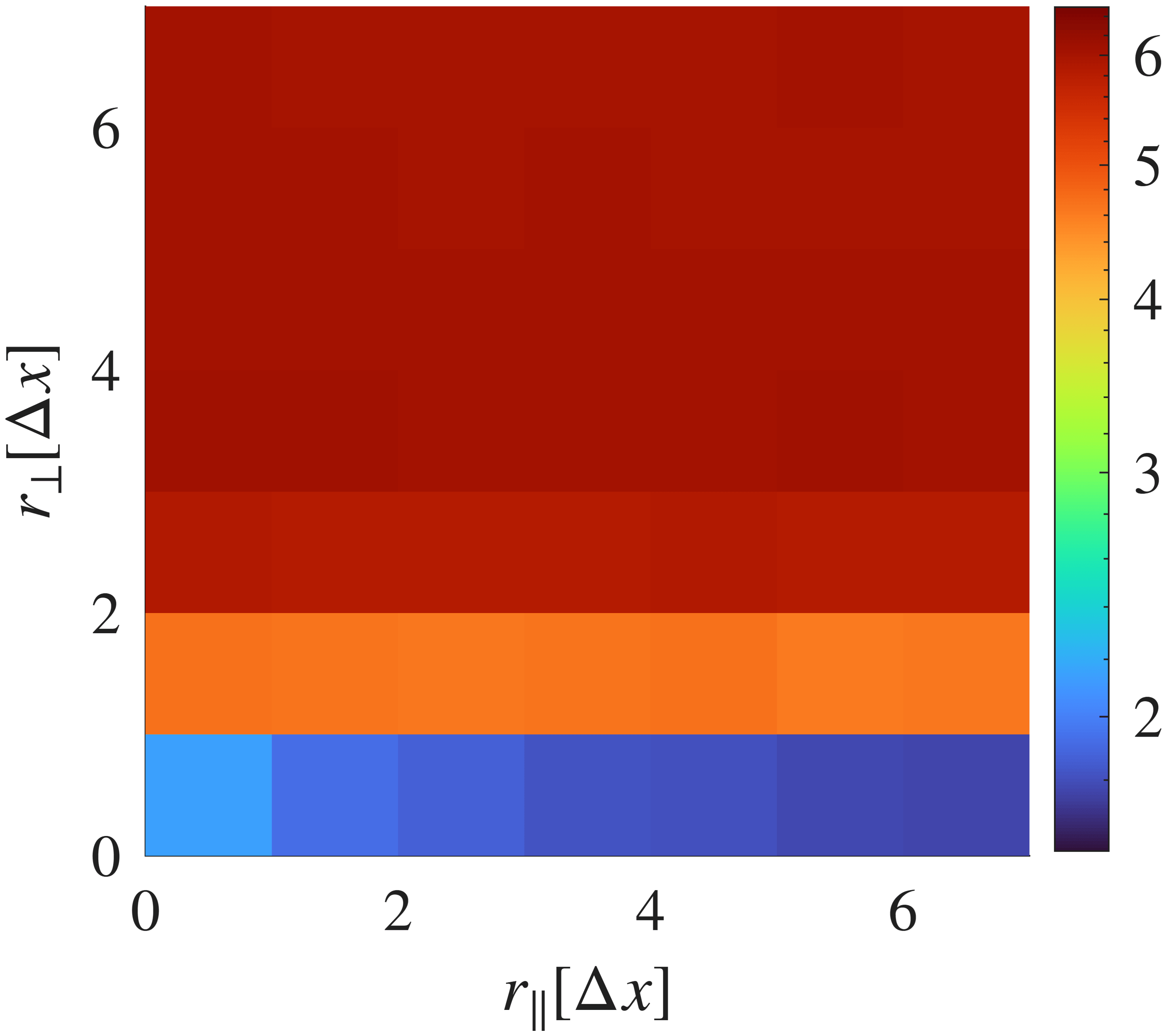}
    \caption{Normalized density structure function (Equation~\ref{equ::den_struct_func}) for the non-linear evolution of the thermal instability in 3D. The color in logarithmic scale indicates the ratio to the overall density variance. The horizontal and vertical axes are in units of cell size, $\Delta x = \tau_\text{syn} c / 256$.}
    \label{fig::thermal_ins_struct_func}
\end{figure}
In this section, we examine the thermal instability driven by synchrotron cooling in the absence of turbulence, as originally proposed by \citet{1967ApJ...150..105S, 1979ApJ...233..463E}. We briefly summarize the underlying physical picture, followed by numerical MHD simulations that track the instability from the linear regime into the nonlinear phase.

The thermal instability amplifies perturbations in density and/or pressure away from the equilibrium state. \citet{1967ApJ...150..105S, 1979ApJ...233..463E, 1990MNRAS.244..530B, 1992A&A...256..689B} examined a uniform, static plasma of density $\rho_0$ and pressure $P_0$, threaded by a uniform magnetic field $B_0 \hat{y}$, where uniform heating $\mathcal{H}$ balances synchrotron cooling in the initial equilibrium,
\begin{equation}
    \mathcal{H}  = \tau_\text{syn}^{-1} P_0 \frac{B_0^2}{\rho_0 c^2} \frac{K_3 \left(\rho_0 c^2 / P_{0} \right)}{K_2 \left(\rho_0 c^2 / P_{0} \right)}. \label{equ::equilibrium}
\end{equation}
Throughout this section, we adopt $\tau_\text{syn} \equiv (\eta_\text{syn} c / L)^{-1}$ (Equation~\ref{equ::cooling}) as the timescale unit.

A small perturbation first produces converging (diverging) flows. 
When the wavevector of the perturbation is aligned with the magnetic field, it induces only density/temperature variations: the cooling rate decreases (increases) in under-dense (over-dense) regions, and the resulting pressure gradient drives a compensating flow that restores equilibrium. In contrast, a perturbation perpendicular to the magnetic field, as shown in Figure~\ref{fig::cool_instability_carton}, advects the field into over-dense (under-dense) regions. The cooling rate therefore increases (decreases) in over-dense (under-dense) regions, satisfying the isobaric thermal-instability criterion \citep{1965ApJ...142..531F},
\begin{equation}
    \left(\frac{\partial P_\text{syn}}{\partial \Theta_t}\right)_{P_g} < 0, \quad \text{if} \quad B \propto \rho, \label{equ::field}
\end{equation}
and the induced pressure gradient further enhances the converging (diverging) flow. This positive feedback amplifies the perturbation, ultimately producing dense, cold regions and dilute, hot regions perpendicular to the magnetic field. \citet{1967ApJ...150..105S, 1979ApJ...233..463E} derived the corresponding linear growth rate: it increases with wavenumber $k$ and saturates at large $k$, with the maximum rate governed by the initial ratio $B_0^2/P_0$ and the cooling timescale $\left(\partial P_\text{syn}/\partial P_g\right)^{-1}$ (see Figure~2 of \citealp{1967ApJ...150..105S}). 

We begin with a 1D single-mode simulation similar to \citet{1990MNRAS.244..530B, 1992A&A...256..689B}, which also serves as a benchmark test for the synchrotron cooling term (Equation~\ref{equ::cooling}). An eigenmode perturbation is applied to the density $\delta\rho$, gas pressure $\delta P_g$, and magnetic field $\delta B_y$, perpendicular to the background field $B_0 \hat{y}$. The initial background plasma is static with $\rho_0$, $P_0 = 2\rho_0 c^2$, and $B_0$ satisfying $B_0^2 = \rho_0 c^2$. The perturbation is sinusoidal with amplitude $\delta\rho/\rho_0 = 10^{-5}$ and wavelength $\lambda = \tau_\text{syn} c$, resolved by $10^4$ cells in a periodic domain. For $t \lesssim 40\, \tau_\text{syn}$, the mode grows exponentially in the linear regime (top panel of Figure~\ref{fig::cool_instability_single}), retaining its sinusoidal form: density and magnetic-field perturbations remain in phase, while $\delta P_g$ is out of phase. Once the fractional amplitude reaches $\sim 0.1$, nonlinear effects emerge (middle panel), where the enhanced $B^2/P_g$ ratio in over-dense regions slows down the instability. By $\sim 58\, \tau_\text{syn}$, the instability produces a cold dense region and a hot dilute region separated by sharp discontinuities (bottom panel). The evolution qualitatively agrees with previous 1D simulations \citep{1990MNRAS.244..530B, 1992A&A...256..689B}.

We then perform 3D simulations to study the nonlinear evolution across all wavenumbers. The plasma is initialized as uniform and static, with small random pressure perturbations $\delta P_g / P_0 \sim \mathcal{U}(-10^{-5}, 10^{-5})$, which trigger the thermal instability in all directions. The 3D periodic domain has size $L^3 = (\tau_\text{syn} c)^3$, resolved by $256^3$ cells. Cold dense phases and hot dilute ones rapidly develop from short-wavelength modes, reaching the nonlinear stage by $\sim 37\, \tau_\text{syn}$. Using the density structure function, we find that density fluctuations grow preferentially with wavevectors perpendicular to the magnetic field, forming field-aligned filaments with transverse scales of a few cells (a few $\Delta x$), reflecting the dominance of high-wavenumber modes.

At the end of this section, we mention several numerical challenges in simulating the thermal instability induced by synchrotron cooling. First, with uniform volumetric heating, the instability is always dominated by the highest-wavenumber modes, whose wavelengths are limited by numerical resolution (Figure~\ref{fig::thermal_ins_struct_func}), preventing convergence in the nonlinear stage. Second, as the plasma cools down or gets evacuated, pressure and density eventually reach the numerical floors, as seen after $58\, \tau_\text{syn}$ in the 1D simulation (Figure~\ref{fig::cool_instability_single}) and after $37\, \tau_\text{syn}$ in 3D (Figure~\ref{fig::thermal_ins_struct_func}). 
Finally, the synchrotron cooling term (Equation~\ref{equ::cooling}) is implemented with a first-order IMEX scheme, so only high-resolution simulations can accurately capture the linear growth stage; however, the temperature and density discontinuities (Figure~\ref{fig::cool_instability_single}) remain sensitive to numerical resolution and affect the non-linear evolution.

\section{Dependence on initial magnetization and numerical resolution}
\label{sec::mag}

\begin{figure}
    \centering
    \includegraphics[width=0.45\textwidth]{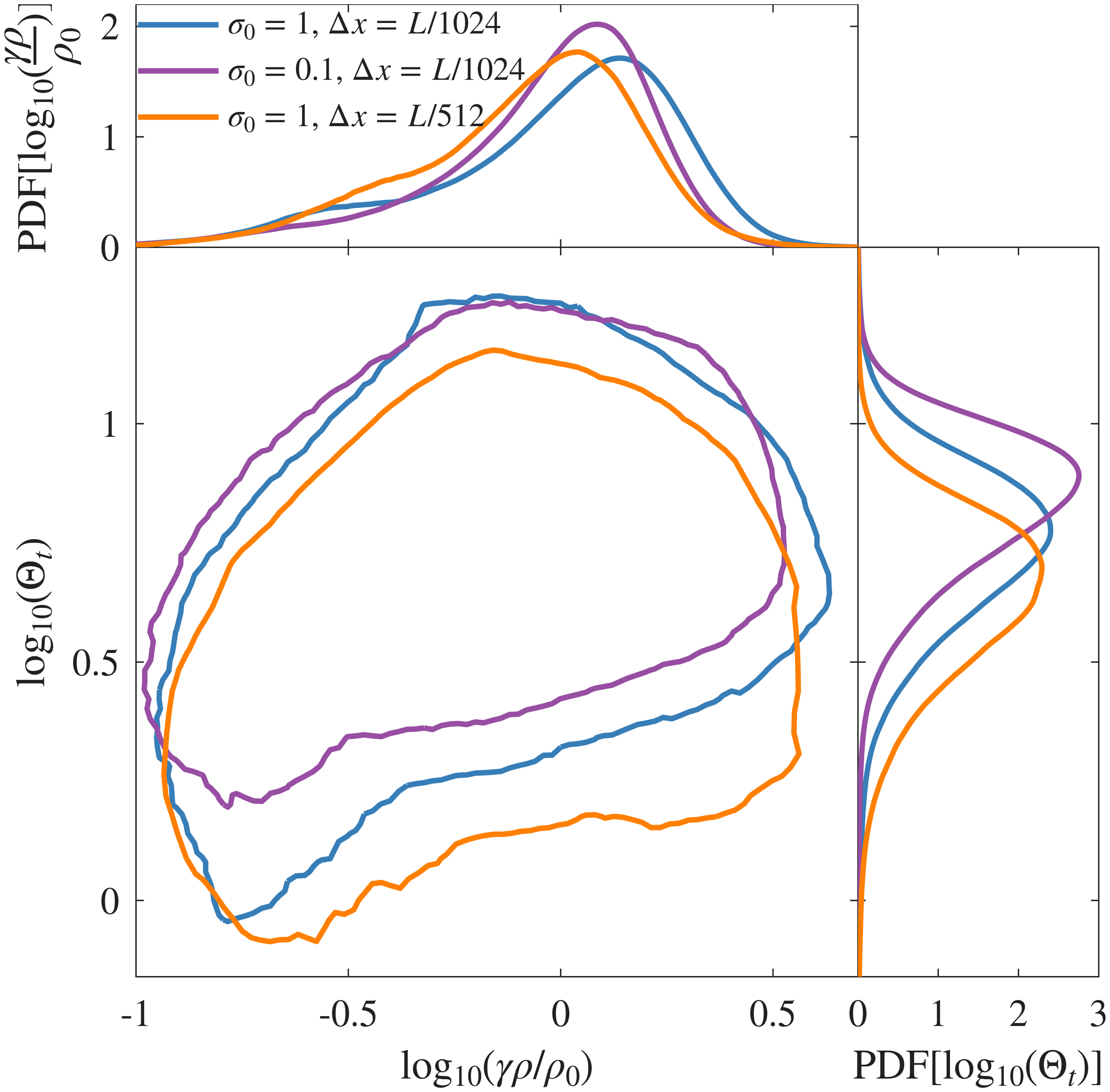}
    \caption{Phase distributions in quasi-steady state for runs with different initial magnetization $\sigma_0$ or different numerical resolution $\Delta x$ but with the same synchrotron cooling efficiency $\eta_\text{syn} = 6.4 \times 10^{-3}$. Colors indicate different runs as shown in the legend.}
    \label{fig::phase_mag}
\end{figure}

\begin{figure}
    \centering
    \includegraphics[width=0.45\textwidth]{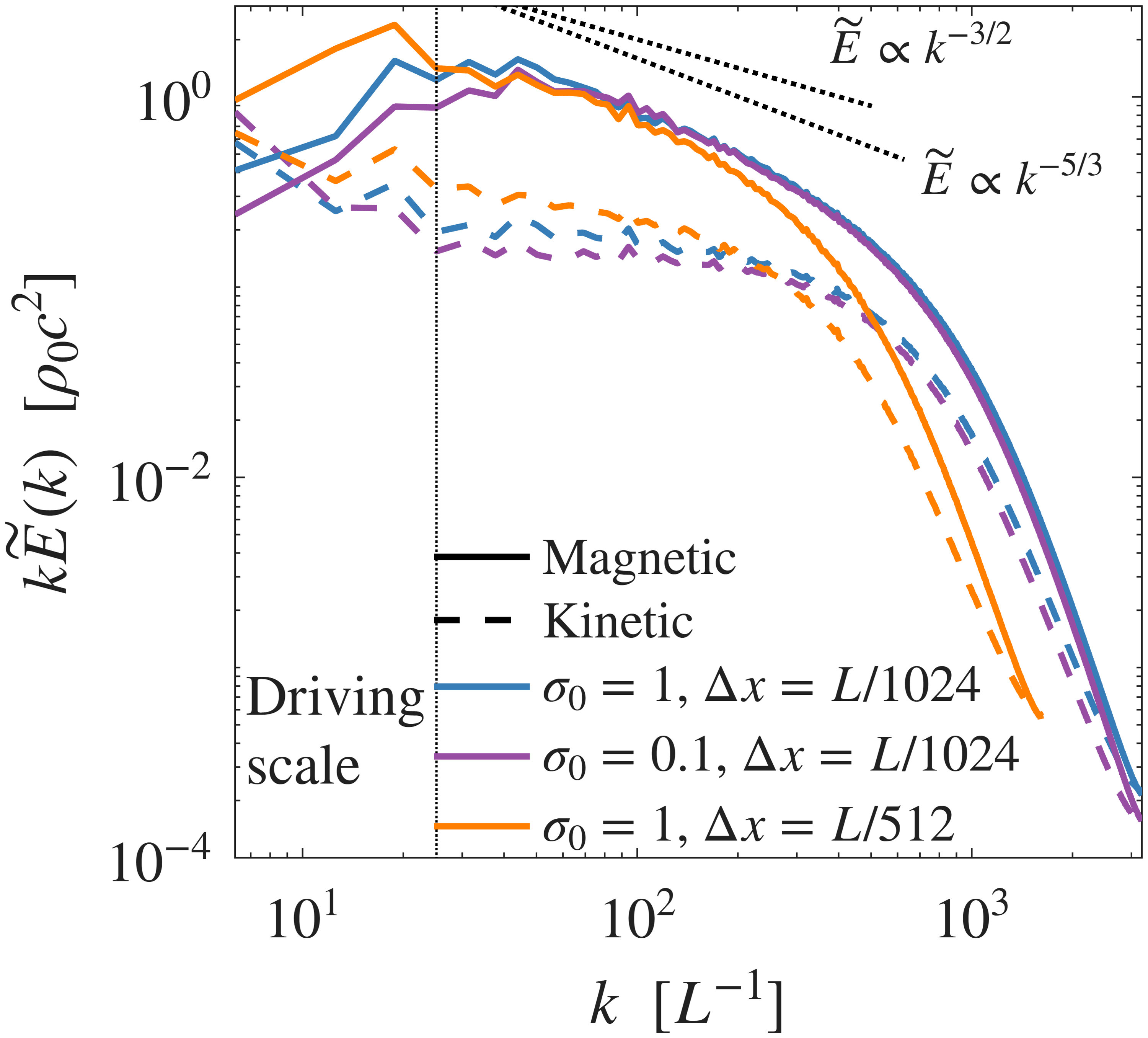}
    \caption{Magnetic (solid lines) and kinetic (dashed lines) power spectral densities in quasi-steady state for two runs with different initial magnetization $\sigma_0$ or different numerical resolution $\Delta x$ but with the same synchrotron cooling efficiency $\eta_\text{syn} = 6.4 \times 10^{-3}$.  Colors indicate different runs as shown in the legend.}
    \label{fig::pwr_enthalpy_mag}
\end{figure}
\begin{figure}
    \centering
    \includegraphics[width=0.45\textwidth]{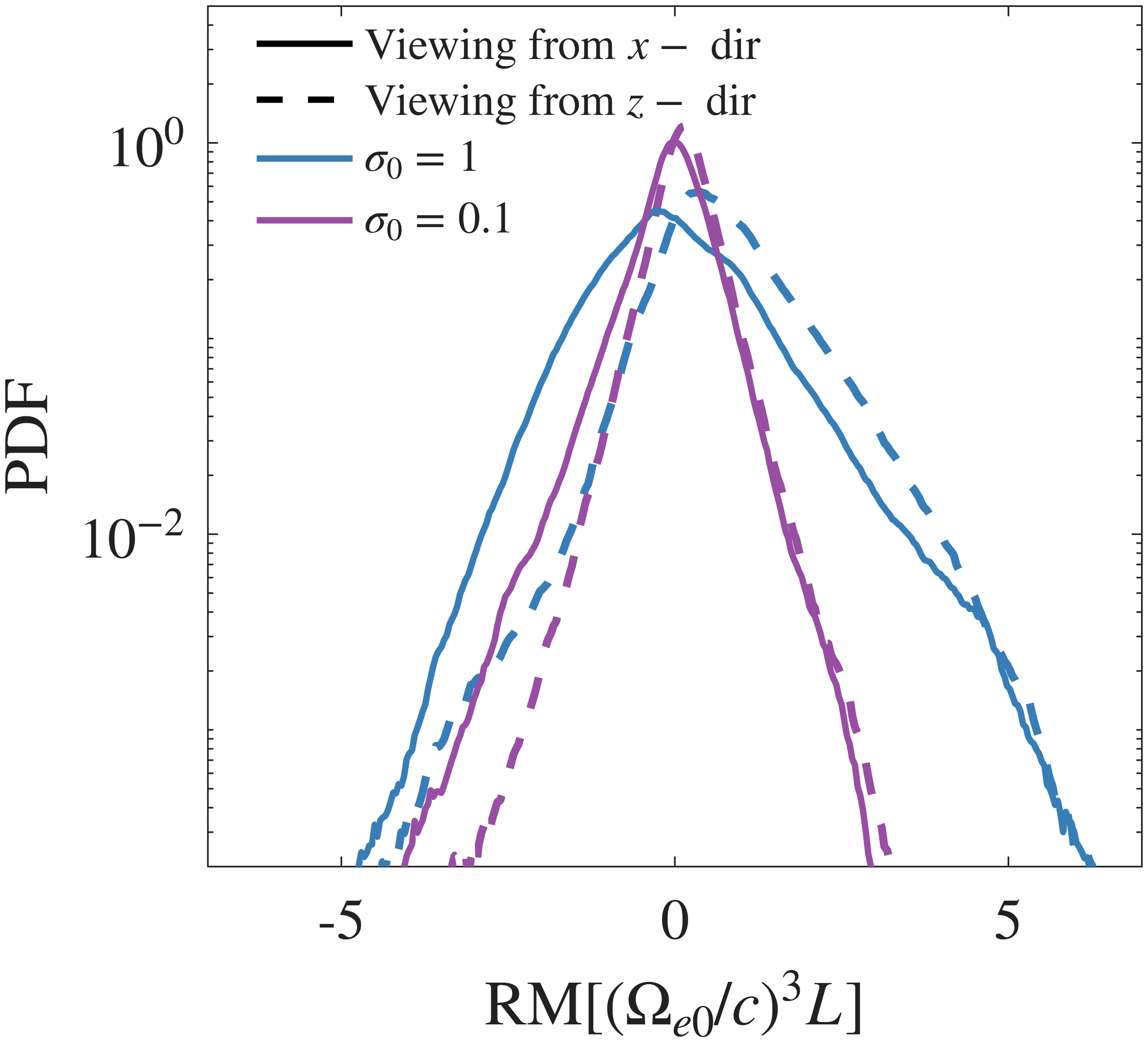}
    \caption{Rotation measure (RM) probability distribution functions (PDFs) in quasi-steady state for two runs with different initial magnetization $\sigma_0$ but with the same synchrotron cooling efficiency $\eta_\text{syn} = 6.4 \times 10^{-3}$. Colors indicate different runs as shown in the legend.}
    \label{fig::rm_pdf_mag}
\end{figure}
We perform two additional simulations to test robustness against initial magnetization and numerical resolution. Fixing $\eta_\text{syn} = 6.4 \times 10^{-3}$ and all the other parameters as in Section~\ref{sec::numerical_setup}, we consider one run with a reduced magnetization of $\sigma_0 = 0.1$ and another with a lower resolution of $\Delta x = L/512$.

The plasma phase distribution is qualitatively the same with varying resolution and initial magnetization (Figure~\ref{fig::phase_mag}). Although minor shifts occur with changes in $\Delta x$ or $\sigma_0$, the overall structure and variance remain similar. These differences are much smaller than those induced by varying $\eta_\text{syn}$ (Figure~\ref{fig::phase}).

Figure~\ref{fig::pwr_enthalpy_mag} compares the magnetic and kinetic power spectral densities (PSDs) in quasi-steady state. The lower-resolution run ($\Delta x = L/512$) exhibits a shorter inertial range due to enhanced numerical dissipation, though its magnetic PSD overlaps with the higher-resolution case ($\Delta x = L/1024$) within the inertial range, while its kinetic PSD is slightly higher. Reducing the initial magnetization $\sigma_0$ yields magnetic and kinetic PSDs that largely overlap with the fiducial run.

The qualitative convergence with varying numerical resolution indicates that our main results are physically robust. The weak dependence on the initial magnetization $\sigma_0$ arises because $\delta B \gg B_0$ in the quasi-steady state, so turbulence is governed by the fluctuating magnetic energy rather than the initial mean field energy. Similarly, other diagnostics, including synchrotron emission and polarization (Sections~\ref{sec::emission}~\&~\ref{sec::syn_stat}), remain similar when varying $\Delta x$ and $\sigma_0$.

In contrast, the $\text{RM}_z$ statistics vary with $\sigma_0$, as shown in Figure~\ref{fig::rm_pdf_mag}: reducing the mean field strength yields a more symmetric distribution around zero, highlighting the mean-field contribution to the rotation measure.
\section{Kinetic power spectra}
\label{sec::kin}
\begin{figure*}
    \centering
    \includegraphics[width=\textwidth]{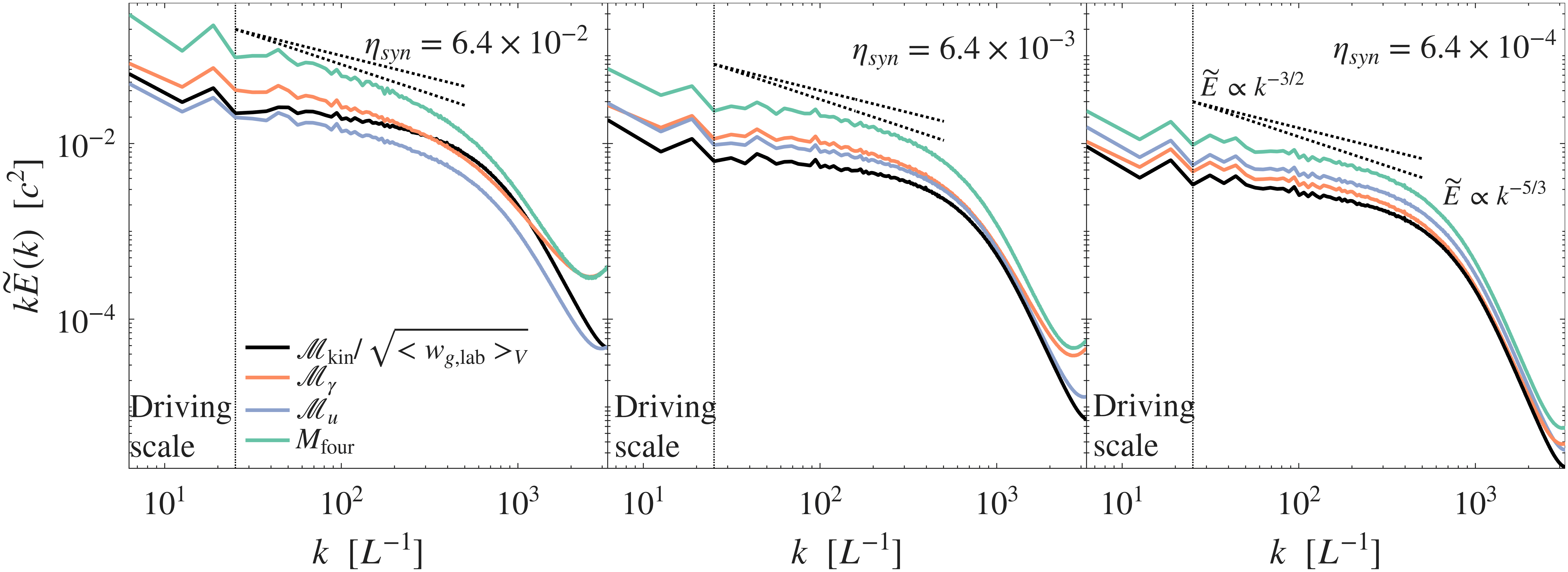}
    \caption{Kinetic power spectral densities (PSDs) in quasi-steady state computed using different definitions of the kinetic field: $\boldsymbol{\mathcal{M}}_\text{kin}$ (black; Equation~\ref{equ::kin_pwr_def}), $\boldsymbol{\mathcal{M}}_\gamma$ (orange; definition \ref{equ::fake_pwr_def}), $\boldsymbol{\mathcal{M}}_u$ (magenta; definition \ref{equ::three_pwr_def}), and $\mathbb{M}_\text{four}$ (viridis; definition \ref{equ::four_pwr_def}). Note that in this figure, colors indicate different definitions of the kinetic field, while panels from left to right correspond to different $\eta_\text{syn}$ from high to low. Two black dotted lines show reference slopes of $\tilde{E}\propto k^{-5/3}$ and $k^{-3/2}$.}
    \label{fig::pwr_kin_compare}
\end{figure*}
The definition of kinetic power spectral density (PSD) in relativistic MHD remains ambiguous. The Newtonian kinetic PSD was originally formulated for incompressible velocity fluctuations \citep{1941DoSSR..30..301K}. Here we consider several velocity-based alternatives:
\begin{enumerate}
    \item $\boldsymbol{\mathcal{M}}_\gamma(\boldsymbol{x}) \equiv \sqrt{\dfrac{1}{\gamma(\boldsymbol{x})+1}} \gamma\boldsymbol{u}(\boldsymbol{x})$: motivated by Equation~\ref{equ::kin_pwr_def} but omitting the enthalpy term; \label{equ::fake_pwr_def}
    \item $\boldsymbol{\mathcal{M}}_u(\boldsymbol{x}) \equiv \boldsymbol{u}(\boldsymbol{x})$ \citep{2013ApJ...766L..10R}; \label{equ::three_pwr_def}
    \item $\mathbb{M}_\text{four}(\boldsymbol{x}) \equiv (\gamma, \gamma\boldsymbol{u})$ \citep{2013ApJ...763L..12Z}. \label{equ::four_pwr_def}
\end{enumerate}
For definitions \ref{equ::fake_pwr_def} and \ref{equ::three_pwr_def}, the kinetic PSDs are obtained from the squared modulus of the Fourier transform of the vector field (as in Equation~\ref{equ::kin_pwr_def}). Definition \ref{equ::four_pwr_def} treats the kinetic field as a four-vector and computes the PSD from the inner product of its Fourier transform.

We compare the kinetic PSDs obtained from these definitions in quasi-steady state in Figure~\ref{fig::pwr_kin_compare} for three runs with different synchrotron cooling efficiencies $\eta_\text{syn}$ / different levels of steady-state enthalpy density (see Figure~\ref{fig::time}). To place all spectra in consistent physical units, the PSD computed from $\boldsymbol{\mathcal{M}}_\text{kin}$ (Equation~\ref{equ::kin_pwr_def}) is re-normalized by the volume-averaged enthalpy $\langle w_\text{lab} \rangle_V$, while preserving its shape.

None of these four kinetic PSD definitions overlaps for the same run, and Definitions~\ref{equ::fake_pwr_def}--\ref{equ::four_pwr_def} yield different slopes across runs, even when the same definition is used. The cooling-efficient case ($\eta_\text{syn}=6.4\times10^{-2}$, leftmost panel in Figure~\ref{fig::pwr_kin_compare}), with the lowest inertia and highest bulk velocities (Figure~\ref{fig::time}), exhibits the steepest spectrum in $\mathbb{M}_\text{four}$ (viridis line), approaching $k^{-5/3}$ over a fraction of the inertial range. By contrast, the cooling-inefficient run ($\eta_\text{syn}=6.4\times10^{-4}$, rightmost panel in Figure~\ref{fig::pwr_kin_compare}) is flatter than $k^{-3/2}$ for the same definition (viridis line). Spectral slopes based on $\boldsymbol{\mathcal{M}}_\gamma$ (orange lines) and $\boldsymbol{\mathcal{M}}_u$ (magenta lines) lie between those based on $\boldsymbol{\mathcal{M}}_\text{kin}$ (black lines) and $\mathbb{M}_\text{four}$  (viridis lines), while preserving the trend: cooling-efficient runs are steepest, and cooling-inefficient runs are flattest.

Comparing $\boldsymbol{\mathcal{M}}_\text{kin}$ (black lines) and $\boldsymbol{\mathcal{M}}_\gamma$ (orange lines) shows that including enthalpy fluctuations yields systematically flatter kinetic PSDs: enthalpy variations amplify small-scale kinetic fluctuations. This effect is strongest in cooling-efficient runs, where the thermal instability produces larger small-scale enthalpy contrasts.

Finally, none of the four kinetic-PSD definitions is fully compatible with relativity. A PSD, as a scalar, should be Lorentz invariant; however, $\boldsymbol{\mathcal{M}}_\text{kin}$ (Equation~\ref{equ::kin_pwr_def}), $\boldsymbol{\mathcal{M}}_\gamma$ (Definition~\ref{equ::fake_pwr_def}), and $\boldsymbol{\mathcal{M}}_u$ (Definition~\ref{equ::three_pwr_def}) are not four-vectors, and thus their spectra are frame dependent. Although $\mathbb{M}_\text{four}$ (Definition~\ref{equ::four_pwr_def}) is a four-vector, $\lvert \mathbb{M}_\text{four}(\boldsymbol{x}) \rvert^2 = 1$ does not reproduce kinetic energy density. Moreover, Lorentz boosts affect spatial intervals, so a PSD based solely on spatial Fourier transforms is inherently frame dependent. A Lorentz-invariant kinetic PSD therefore remains an open problem; possible directions include formulations based on four-dimensional spacetime intervals or Lagrangian tracer statistics.
\section{Phase distribution and synchrotron cooling distribution}
\label{sec::phase}
\begin{figure*}
    \centering
    \includegraphics[width=0.9\textwidth]{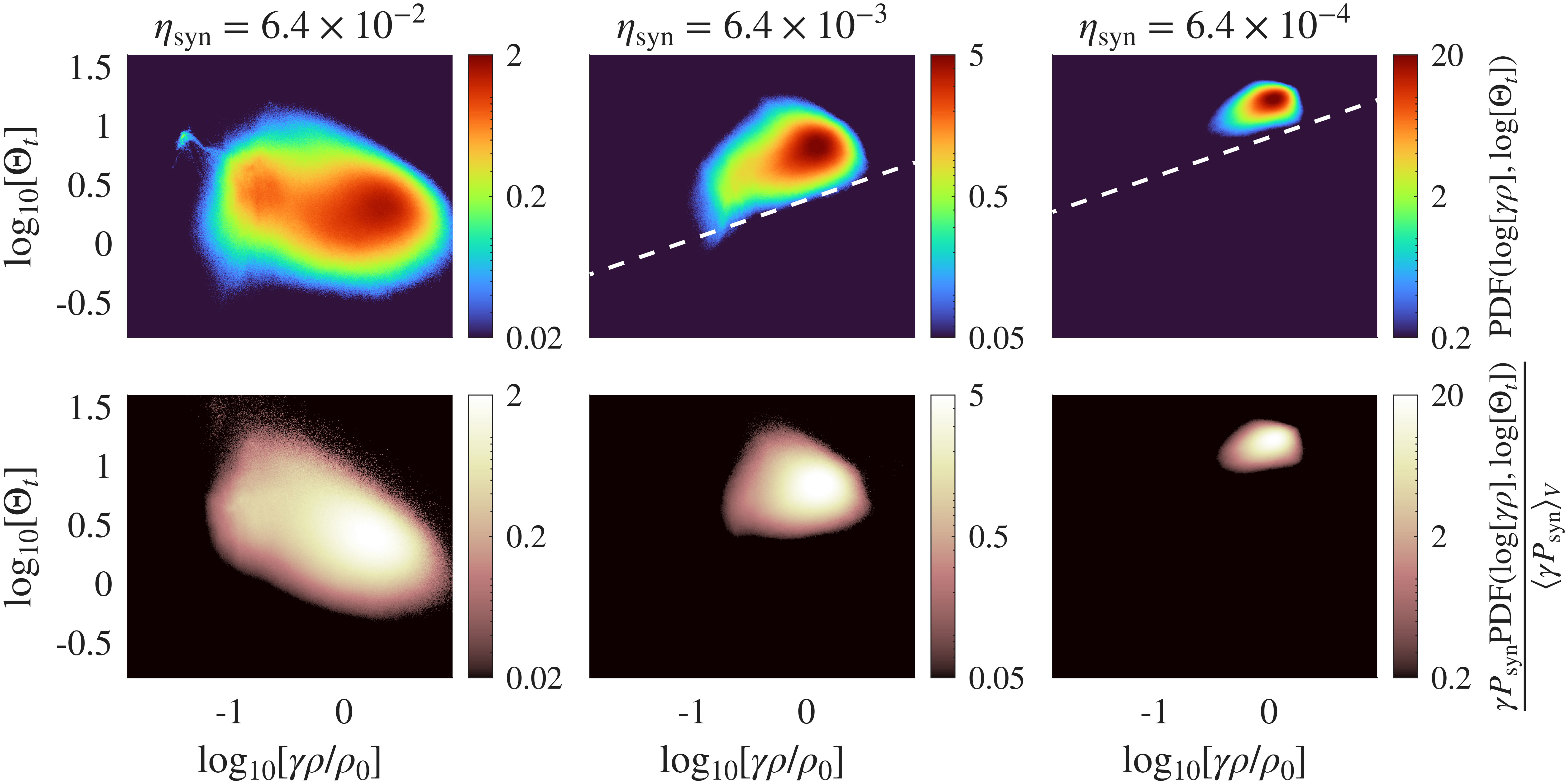}
    \caption{The phase distribution (upper panels) and the synchrotron cooling distribution (lower panels)  in quasi-steady state, for lab-frame density $\gamma \rho$ and  temperature $\Theta_t$, with different synchrotron cooling efficiencies $\eta_\text{syn}$. Upper panels are weighted by volume, lower panels by synchrotron power.
    Note that the colorbars are in logarithmic scale.
    The white dashed lines in the upper panels refer to the adiabatic relation $\Theta_t \propto(\gamma \rho)^{\Gamma - 1}$ with $\Gamma = 4/3$.}
    \label{fig::phase_temp_den}
\end{figure*}
We show the quasi-steady plasma phase distribution for all runs in the upper panels of Figure~\ref{fig::phase_temp_den}, as PDFs in lab-frame density $\gamma\rho$ and temperature $\Theta_t$. As $\eta_\text{syn}$ increases, the distribution broadens and the density-temperature relation shifts from a positive correlation toward an anti-correlation, indicating a stronger role for the thermal instability (Figure~\ref{fig::cool_instability_single}). In the cooling-inefficient run ($\eta_\text{syn}=6.4\times10^{-4}$), the plasma remains nearly uniform, concentrated around $\gamma\rho\sim\rho_0$ and $\Theta_t\sim\langle\Theta_t\rangle_V$ (Figure~\ref{fig::time}), with only modest broadening along the adiabatic scaling $\Theta_t\propto\rho^{\Gamma-1}$ (white dashed lines). In the most cooling-efficient run ($\eta_\text{syn}=6.4\times10^{-2}$), the phase distribution becomes bimodal, with peaks around $(\gamma\rho,\Theta_t)\sim(3\rho_0,2)$ and $(0.2\,\rho_0,3)$, consistent with the outcome of the thermal instability (Section~\ref{sec::thermal_ins}); however, as discussed in the main text, turbulent mixing prevents the sharp phase separation expected in the pure thermal instability scenario.

The lower panels of Figure~\ref{fig::phase_temp_den} show the same PDFs, but weighted by the lab-frame synchrotron cooling power $\gamma P_\text{syn}$. Although Equation~\ref{equ::cooling} implies that cooling is most efficient in high-temperature and high-density regions, the cooling distributions closely follow the phase distributions, only slightly shifting toward higher temperature and/or higher density regions. The spatially inhomogeneous magnetic field plays a key role in shaping the cooling distributions.

\begin{figure*}
    \centering
    \includegraphics[width=0.9\textwidth]{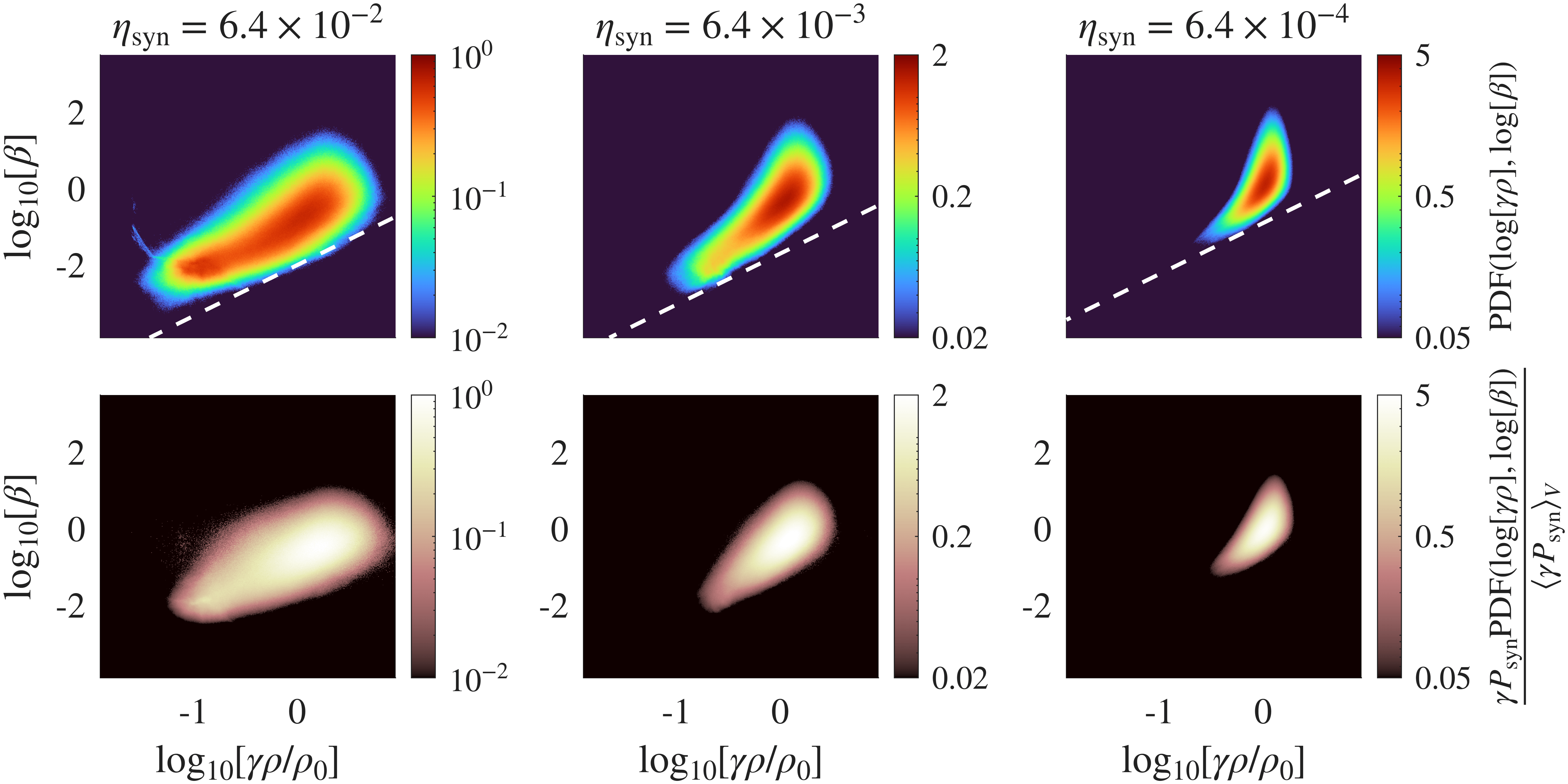}
    \caption{The distribution of plasma $\beta \equiv 2 P_g/B'^2$ and lab-frame density $\gamma \rho$ in quasi-steady state, with different synchrotron cooling efficiencies $\eta_\text{syn}$. Upper panels are weighted by volume, lower panels by synchrotron power.
    Note that the colorbars are in logarithmic scale.
    The white dashed lines refer to the adiabatic relation $P_g \propto \left(\gamma \rho\right)^{\Gamma }$ with $\Gamma = 4/3$.}
    \label{fig::phase_beta_den}
\end{figure*}

To understand the relation between thermodynamic state and magnetic field strength, Figure~\ref{fig::phase_beta_den} shows the distributions of plasma $\beta \equiv P_g/B'^2$ and lab-frame density $\gamma\rho$. In the cooling-inefficient run ($\eta_\text{syn}=6.4\times10^{-4}$), which largely follows the adiabatic scaling $P_g\propto(\gamma\rho)^\Gamma$ (white dashed lines), the $\beta$-$\gamma\rho$ relation becomes steeper than the while dashed line due to weaker magnetic fields in high-pressure regions and vice versa. Similar anti-correlations between magnetic field strength and gas pressure have been observed in non-relativistic turbulence \citep{2017ApJ...836...69Y,2016ApJ...831...85Y,2020ApJ...889...19G}, possibly due to pressure balance constraints or magnetosonic waves \citep{2003A&A...398..845P}. This anti-correlation weakens as $\eta_\text{syn}$ increases, reflecting the growing influence of the thermal instability. However, turbulent compression still dominates, because the $\beta$-$\gamma\rho$ relation does not fully reverse, as would be expected for pure thermal instability (Figure~\ref{fig::cool_instability_single}).

In summary, the phase structure is jointly shaped by turbulent motions and thermal instability. As the cooling efficiency increases, thermal instability becomes more important, while turbulent mixing prevents the formation of a sharply separated multiphase medium.

\section{Structure functions and anisotropy}
\label{sec::struct_func}
\begin{figure}
    \centering
    \includegraphics[width=0.45\textwidth]{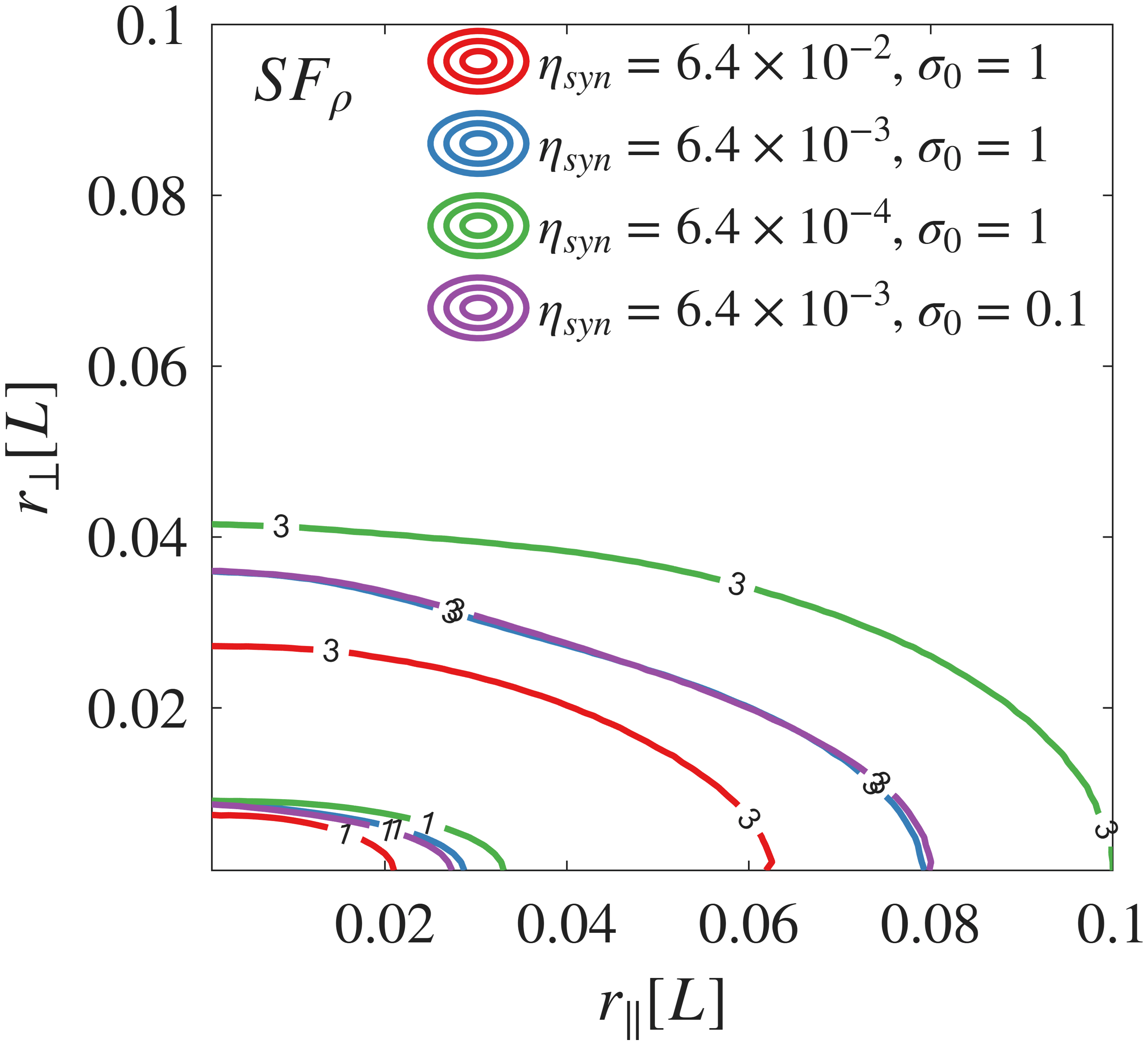}
    \caption{Normalized density structure function of quasi-steady turbulence. Colors indicate different cooling efficiencies and initial magnetization as shown in the legend. The number over each line indicates the ratio to the overall density variance (Equation~\ref{equ::den_struct_func}).}
    \label{fig::den_struct_func}
\end{figure}

We compute second-order structure functions of lab-frame density $\rho_{\rm lab}\equiv\gamma\rho$ and magnetic field in quasi-steady state. 
The normalized density structure function is computed as
\begin{align}
    SF_\rho \left(r_\parallel, r_\perp\right) &=
    \frac{\left\langle\left(\rho_{\rm lab}(\boldsymbol{x}+\boldsymbol{r})-\rho_{\rm lab}(\boldsymbol{x})\right)^2\right\rangle_V}{\left\langle\left(\rho_{\rm lab}(\boldsymbol{x})-\langle\rho_{\rm lab}\rangle_V\right)^2\right\rangle_V}, \label{equ::den_struct_func} \\
    \text{with } r_\parallel &= \boldsymbol{r}\cdot\frac{\boldsymbol{B}(\boldsymbol{x}+\boldsymbol{r})+\boldsymbol{B}(\boldsymbol{x})}{|\boldsymbol{B}(\boldsymbol{x}+\boldsymbol{r})+\boldsymbol{B}(\boldsymbol{x})|},\quad r_\perp=\sqrt{|\boldsymbol{r}|^2-r_\parallel^2}. \notag
\end{align}
Here $\boldsymbol{r}$ is decomposed into components parallel and perpendicular to the local field, and the structure function is normalized by the density variance to facilitate` comparison across runs. 
Figure~\ref{fig::den_struct_func} shows $SF_\rho$ for all runs: the structure function is clearly elongated along the field; the cooling-efficient turbulence produces sharper transverse contrasts, while the cooling-inefficient run is smoother. Runs with the same $\eta_\text{syn}$ but different $\sigma_0$ yield nearly identical $SF_\rho$.

\begin{figure}
    \centering
    \includegraphics[width=0.45\textwidth]{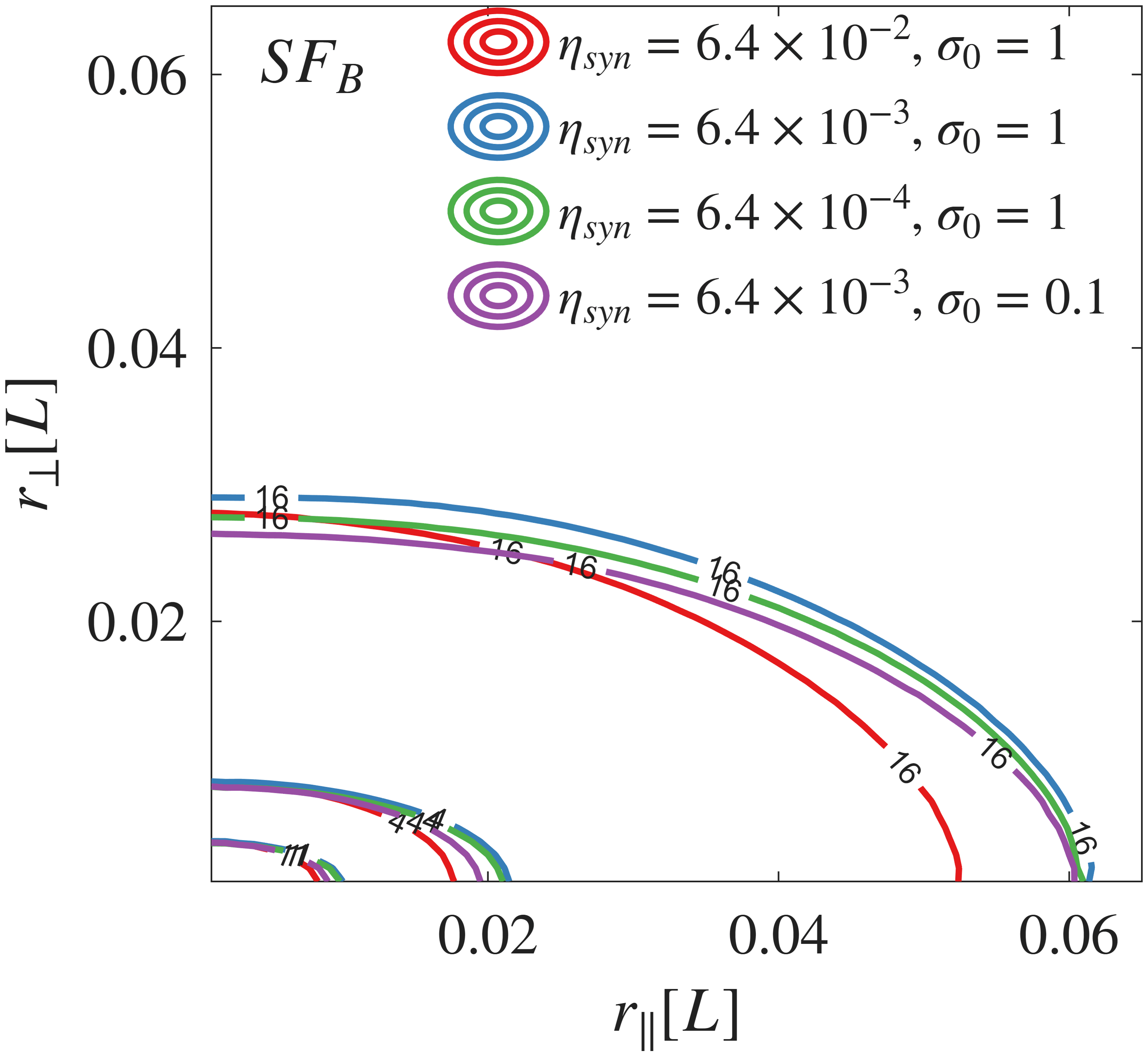}
    \caption{Magnetic structure function of quasi-steady turbulence. Colors indicate different cooling efficiencies and initial magnetization as shown in the legend. The number over each line indicates the magnitude of the structure function (Equation~\ref{equ::mag_struct_func}).}
    \label{fig::mag_struct_func}
\end{figure}
We also show the magnetic structure function in Figure~\ref{fig::mag_struct_func} \citep{2009ApJ...701..236C},
\begin{align}
    SF_B\left(r_\parallel, r_\perp\right)=\left\langle\left|\boldsymbol{B}(\boldsymbol{x}+\boldsymbol{r})-\boldsymbol{B}(\boldsymbol{x})\right|^2\right\rangle_V, \label{equ::mag_struct_func}
\end{align}
Here $SF_B$ is left un-normalized because the total magnetic energy is similar across runs (Figure~\ref{fig::time}). As with density, $SF_B$ is elongated along the local magnetic field and shows no substantial dependence on $\eta_\text{syn}$ or $\sigma_0$.

\begin{figure}
    \centering
    \includegraphics[width=0.45\textwidth]{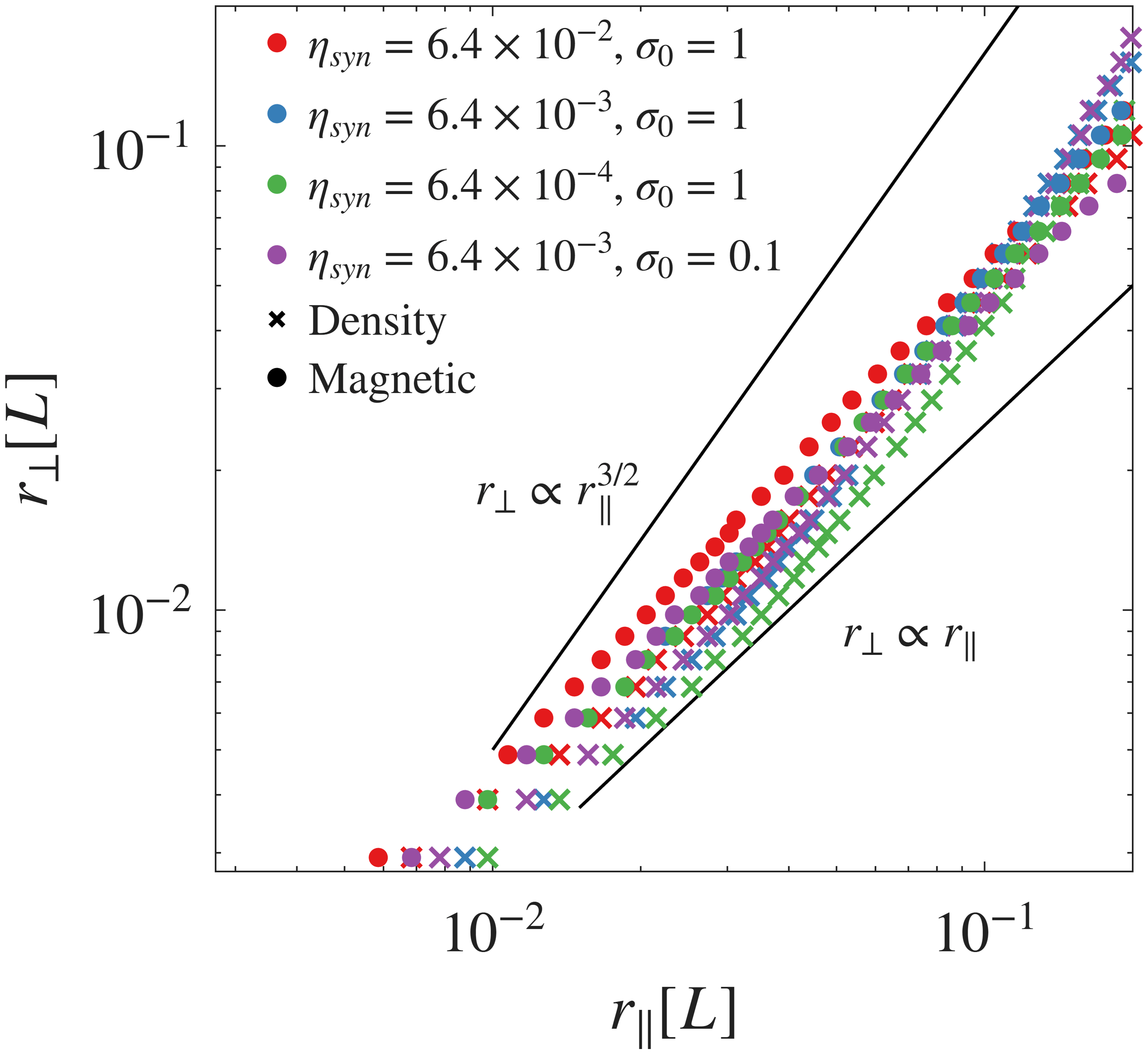}
    \caption{Anisotropy in density (crosses, computed from Figure~\ref{fig::den_struct_func}) and magnetic (dots, computed from Figure~\ref{fig::mag_struct_func}) structure functions. Colors indicate different cooling efficiencies and initial magnetization as shown in the legend. The solid lines represent the Goldreich-Sridhar scaling $r_\perp \propto r_\parallel^{3/2}$ \citep{1995ApJ...438..763G} and the isotropic scaling $r_\perp \propto r_\parallel$.}
    \label{fig::aniso}
\end{figure}
The anisotropy of the density and magnetic structure functions (Figures~\ref{fig::den_struct_func} and \ref{fig::mag_struct_func}) is summarized in Figure~\ref{fig::aniso} by comparing $r_\parallel$ and $r_\perp$ at fixed structure-function values (the normalization does not affect this measure). Anisotropy develops from large to small scales following the Goldreich-Sridhar scaling $r_\parallel \propto r_\perp^{2/3}$ \citep{1995ApJ...438..763G}. At the smallest scales, dominated by dissipation, the anisotropy ratio $r_\parallel/r_\perp$ becomes constant. We find no significant dependence of the anisotropy on the initial magnetization or the cooling efficiency, but we note that  density exhibits stronger anisotropy than magnetic field, likely a consequence of synchrotron-cooling-induced thermal instability (Section~\ref{sec::cooling_in_turb} \& Appendix~\ref{sec::thermal_ins}).

\bibliographystyle{aasjournal}
\bibliography{citation}
\end{document}